\documentclass[pra, notitlepage, showpacs, floatfix, reprint, citeautoscript, superscriptaddress]{revtex4-1}
\usepackage{header}

\begin{document}

\title{Phonon Hall viscosity from phonon-spinon interactions}

\author{Yunchao Zhang}
\affiliation{Department of Physics, Harvard University, Cambridge MA 02138, USA}
\author{Yanting Teng}
\affiliation{Department of Physics, Harvard University, Cambridge MA 02138, USA}
\author{Rhine Samajdar}
\affiliation{Department of Physics, Harvard University, Cambridge MA 02138, USA}
\author{Subir Sachdev}
\affiliation{Department of Physics, Harvard University, Cambridge MA 02138, USA}
\author{Mathias S.~Scheurer}
\affiliation{Institute for Theoretical Physics, University of Innsbruck, Innsbruck A-6020, Austria}

\begin{abstract}
Motivated by experimental observations, Samajdar {\it et al.\/} [Nature Physics {\bf 15}, 1290 (2019)] have proposed that the insulating N\'eel state in the parent compounds of the cuprates is proximate to a quantum phase transition to a state in which N\'eel order coexists with semion topological order. We study the manner in which proximity to this transition can make the phonons chiral, by inducing a significant phonon Hall viscosity. We describe the spinon-phonon coupling in a lattice spinon model coupled to a strain field, and also using a general continuum theory constrained only by symmetry. We find a nonanalytic Hall viscosity across the transition, with a divergent second derivative at zero temperature.
\end{abstract}

\maketitle
\section{Introduction}
Quantum spin liquids (QSLs) are exotic phases of matter arising from highly correlated spins with frustrated interactions, in which zero-point fluctuations are so strong that spin ordering is prevented even down to zero temperature \cite{anderson1987resonating, kivelson1987topology}. QSLs often host a wide variety of collective phenomena, including topological degeneracy and long-range entanglement \cite{wen2002,balents2010spin,savary2016quantum, lee2008end} that make them ideal for theoretical study. Most remarkably, QSLs are characterized by nonlocal fractionalized excitations \cite{senthil2001fractionalization}, such as charge-neutral ``spinons'' coupled to emergent gauge fields \cite{sachdev2018topological}. The spinons can be either gapped or gapless and may be bosons or fermions, depending on the scenario \cite{read1991large, senthil2000z}. 

While there have been extensive experimental efforts towards detection of QSLs, unambiguous evidence remains elusive \cite{broholm2020quantum}. The measurement of topological properties of QSLs is difficult since unlike conventional quasiparticles, spinons are invisible to local probes. One current line of thinking therefore aims to study QSLs by looking for signatures of fractionalization through the interactions of spinons with other degrees of freedom in the system. 

In this paper, we will study the coupling of spinons to lattice excitations. Specifically, we will be interested in a response coefficient called the Hall viscosity \cite{avron95,avron98}. Similar to the Hall conductance, the phonon Hall viscosity can appear for phonons coupled to a gapped electronic system that breaks time-reversal symmetry. The Hall viscosity tensor $\eta_{ijkl}$ characterizes the system's viscoelastic response to a strain deformation as
\begin{align}
    \ev{\pdv{H}{\epsilon^{}_{ij}}} = \lambda^{}_{ijkl} \epsilon^{}_{kl} + \eta^{}_{ijkl} \dot{\epsilon}^{}_{kl},
\end{align}
where $\epsilon_{ij} \equiv (\partial_i u_j + \partial_j u_i)/2$ is the symmetrized strain tensor, and the time derivative is represented by the overdot. In the presence of $C_4$ symmetry in two dimensions ($2$D), there is only one independent component of the Hall viscosity tensor, $\eta_{xxxy}$ \cite{lifshitz1986}. Contrary to a viscosity that is dissipative, the Hall viscosity is antisymmetric with respect to the pairs of indices $(ij)$ and $(kl)$ and hence, nondissipative \cite{dissipation_note}. 

The Hall viscosity was first studied in the context of the quantum Hall effect, in which it was shown to be proportional to the square of the electron filling density for integer quantum Hall fluids \cite{avron98,read2009,read2011,bradlyn2012,son2012,barkeshli2012,abanov2017,haldane2014,parrikar2014,hughesfradkin2011,hughes2013long}. In these systems, the Hall viscosity can be calculated as the response of an appropriate continuum field theory to a variation of the underlying geometry or spatial metric, $g_{ij}$. This Hall viscosity, which acts as a Chern-Simons-type term for the frame field, was termed the \textit{gravitational} Hall viscosity in Ref.~\onlinecite{barkeshli2012}. Instead, our focus will be on the response of systems of phonons and the resulting \textit{phonon} Hall viscosity. While the Hall viscosity originates from the chiral spinons, we use the term \textit{phonon} Hall viscosity to indicate that the stress tensor is coupled to the lattice acoustic phonons instead of a background spatial metric. This is appropriate because the resulting equations of motion for the phonons have a corresponding Hall viscosity term.

Theoretically, the phonon Hall viscosity has been studied for electronic systems and topological insulators \cite{barkeshli2012, shapourian2015viscoelastic, son2012,shitade2014, cho2014geometry}. For lattice systems, such as discrete tight-binding models, there are \textit{a priori} many different ways to model the viscoelastic response, including coupling to a lattice frame field \cite{shapourian2015viscoelastic} or using momentum polarization methods \cite{tu2013,zaletel2015}. We will adopt a more physical ``geometric bond stretching'' approach, realizing the strain as a modification to the tight-binding overlap integrals originating from the lattice sites being displaced from their equilibrium positions. This coincides with the approach of viewing the phonon Hall viscosity as the adiabatic response of a system to acoustic phonons \cite{barkeshli2012}, analogous to the Hall conductance. Using the Kubo formula, the Hall viscosity can also be recognized as a type of Berry curvature of the ground-state wavefunction.

While a measurement of the Hall viscosity would provide valuable information for identifying phases with topological order, it has been difficult to do so in practice. Nevertheless, it is possible to experimentally detect the phonon Hall viscosity through other physical quantities that share the broken symmetries. For example, one such quantity is the phonon thermal Hall conductivity, which can be nonzero only with broken time-reversal and (in-plane) mirror symmetries. In fact, recent experiments by \citet{grissonnanche2020chiral} and \citet{boulanger2020thermal} suggest that chiral phonons are responsible for the large thermal Hall conductivities measured in the insulating phase of several cuprate superconductors. A nonzero phonon Hall viscosity could be a mechanism for intrinsic phonon chirality in these systems. The phonon Hall viscosity leads to both intrinsic and extrinsic contributions to the thermal Hall conductivity: the intrinsic contribution is discussed in Section~\ref{sec:exp}, while the extrinsic contribution is discussed in Ref.~\onlinecite{haoyu_subir}.

We will study the phonon Hall viscosity induced by lattice strain couplings to a chiral spin liquid on the square lattice. In particular, we are interested in a spin-liquid ansatz in which the orbital coupling of the applied magnetic field drives the conventional confining N\'eel insulator to a state with semion topological order \cite{rhine2019enhanced, wangsenthil2017}. Recent optical experiments by \citet{Hsieh20}
indicate the presence of mirror-plane-symmetry breaking which is compatible with this scenario.
In our paper, we will analyze the behavior of the Hall viscosity in both the lattice tight-binding model and the continuum Dirac field theory. We find that the above mentioned quantum phase transition (QPT) in the spinon sector is reflected by a divergence in the second derivative of the phonon Hall viscosity. 

The rest of the paper is organized as follows. We begin in Sec.~\ref{sec:phonon_HV} by reviewing the general definition of the phonon Hall viscosity and linear response theory. Section~\ref{sec:ansatz} introduces the mean-field chiral spin liquid model on the square lattice. We study the spinon-phonon interactions in two settings. On the lattice, we consider phonon-fermion coupling by ``bond stretching'' in Sec.~\ref{sec:Lattice}, whereas for the continuum field theory, we couple phonons and spinons based on symmetry considerations in Sec.~\ref{sec:continuum}. After commenting on some physical consequences in Sec.~\ref{sec:exp}, we summarize and discuss our results in Sec.~\ref{sec:conclusion}. 

\section{Phonon Hall Viscosity}\label{sec:phonon_HV}

\subsection{Phonon effective action with broken time-reversal symmetry}
For gapped fermionic systems, the low-energy dynamics of acoustic phonons is captured by an effective action for $\vec{u}(\vec{r})$, describing the displacement of an atom from its original location. The effective action obtained by integrating out the fermionic degrees of freedom is
\begin{align}
 \mathcal{Z} &= \int \mathcal{D}\bar{\psi}\, \mathcal{D}\psi \,\mathcal{D}\vec{u}\;e^{-  S(\vec{u},\bar{\psi}, \psi)}= \int \mathcal{D}\vec{u}\; e^{-S_\text{eff}(\vec{u})}.
\end{align}
In the long-wavelength limit, the phonon effective action is determined by the mass density $\rho$ and the elastic moduli tensor $\lambda_{ijkl}$,
\begin{equation}
\label{eq:quadratic_phonon_action}
    S^{}_{\rm eff}=\dfrac{1}{2}\int d^dx\,dt\; \left(\rho \dot{u}^{}_j\dot{u}^{}_j-\lambda^{}_{ijkl}\partial^{}_i u^{}_j\partial^{}_k u^{}_l\right).
\end{equation} 
For gapless states such as metals, the phonon action will generally be nonlocal and thus cannot be written as above. 

When time-reversal symmetry is broken, there is an allowed, nondissipative Hall viscosity term \cite{barkeshli2012, avron95, avron98}
\begin{align}
\label{eq:delta_s_eta}
    \delta S =  \frac{1}{2} \int d^{d} x\, d t\; \eta^{}_{ijkl} \partial^{}_i u^{}_j \partial^{}_k \dot{u}^{}_l,
\end{align} with $\eta_{ijkl}=-\eta_{klij}$ antisymmetric under the exchange of pairs of indices. The number of independent components of $\eta_{ijkl}$ can, in general, be determined  using symmetry. For example, one can show that $\eta_{ijkl}$ will always vanish for a three-dimensional isotropic system. As we are concerned with phonons in a spin-liquid background, we will restrict ourselves to $d=2$ in the subsequent analysis. For simplicity, we will also assume $C_4$-rotation symmetry, though this requirement can easily be relaxed.

Following Ref.~\onlinecite{barkeshli2012}, we can knead the Hall viscosity in Eq.~(\ref{eq:delta_s_eta}) into a more convenient form by defining the strain tensor $\epsilon_{ij}$ and the vorticity (also called rotation) tensor $\theta_{ij}$ according to
\begin{equation}
    \epsilon^{}_{ij} \equiv \frac{1}{2}(\partial^{}_i u^{}_j + \partial^{}_j u^{}_i), \quad \theta^{}_{ij} \equiv \frac{1}{2}(\partial^{}_i u^{}_j - \partial^{}_j u^{}_i). \label{DefinitionOfTensors}
\end{equation}
Dropping boundary terms, Eq.~(\ref{eq:delta_s_eta}) can then be rewritten as
\begin{align}
\delta S&=2 \int d^2{x}\, d{t}\left[\eta^{H}\hspace{-0.3em}\left(\epsilon^{}_{x x}-\epsilon^{}_{y y}\right) \dot{\epsilon}^{}_{x y}+\eta^{M}\hspace{-0.3em}\left(\epsilon^{}_{x x}+\epsilon^{}_{y y}\right) \dot{\theta}^{}_{x y}\right].
\label{eq:Phonon_L_General}
\end{align}
Here, we have defined $\eta^H$\,$=$\,$ (\eta_{xxxy} + \eta_{xxyx})/2$ and $\eta^M = (\eta_{xxxy} - \eta_{xxyx})/2$. While boundary terms can modify surface phonon dispersions for topological insulators and generate interesting effects such as phonon Faraday rotation \cite{shapourian2015viscoelastic, tuegel2017hall}, we will ignore these phenomena in our discussion. Finally, it can also be useful to rewrite the action $\delta S$ in Eq.~\eqref{eq:Phonon_L_General} directly in terms of the deformation field $\vec{u}$. In that case, there ends up being one effective Hall viscosity coefficient $\eta\equiv\eta_{xxxy} = \eta^H + \eta^M$, 
\begin{align}
\delta S= \int d^2{x}\, d{t}\left[\frac{-\eta}{2} \left( \nabla^2 u^{}_x \dot{u}^{}_y -\nabla^2 u^{}_y \dot{u}^{}_x \right)\right]. \label{eq:Intro_hallViscosityLagran}
\end{align}
In the calculations hereafter, however, we will follow Eq.~\eqref{eq:Phonon_L_General} and discuss $\eta^H$ and $\eta^M$ separately.

\subsection{Definition as a response function}
We can also view the Hall viscosity as a response function. To begin, we make the adiabatic assumption that the time scale of the lattice motion is infinitely slower than that of the fermions' motion, so that the electronic configuration is always in its instantaneous ground state with respect to its lattice configuration. This implies that the electronic quasiparticles only couple to phonons that are well below their energy gap. The lattice deformation fields $\vec{u}$ then act to modify the effective hopping terms in the tight-binding Hamiltonian $H_{\rm{t.b.}}$ for the electronic system and can be treated as external parameters. 

In Fourier space, viewing $u(\vec{q}) =\sum_nu({\vec{r}})e^{i  \vec{q}\cdot\vec{r}}/\sqrt{L}$ as parameters in $H_{\rm{t.b.}}$, we can first define the two-component Hall tensor from linear response theory through the Kubo formula \cite{barkeshli2012,bradlyn2012}
\begin{equation}
\eta^{}_{ab}(\vec{q})=\lim_{\omega\rightarrow 0}\dfrac{1}{\omega}\dfrac{1}{L^d}\int d t\;e^{i\omega t}\left\langle \left[\dfrac{\partial H^{}_{\rm t.b.}(t)}{\partial u^{}_{a,\vec{q}}},\dfrac{\partial H^{}_{\rm t.b.}(0)}{\partial u^{}_{b,-\vec{q}}}\right]\right\rangle, 
\end{equation}
where it is clear that the Hall tensor is, by construction, antisymmetric, i.e., $\eta_{ab} = -\eta_{ba}$. 
This leads to an effective action of the form 
\begin{equation}
\label{eq:2componenteta}
\delta S=\dfrac{1}{2}\int \dfrac{d^d{\vec{q}}\,d{t}}{(2\pi)^d}\eta^{}_{a b}(\vec{q})u^{}_{a}(-\vec{q},t)\dot{u}^{}_{b}(\vec{q},t).
\end{equation}
From Eq.~(\ref{eq:2componenteta}), we can obtain the Hall viscosity tensor by taking the appropriate derivatives 
\begin{align}
    \eta^{}_{ijkl} = \frac{1}{2} \lim_{q\rightarrow 0} \pdv{}{{q}^{}_i}\pdv{}{{q}^{}_k}\eta^{}_{jl}(\vec{q}).
\end{align}

\section{Spin Liquid Ansatz on the Square Lattice}
\label{sec:ansatz}

Our model of interest, studied in Ref.~\onlinecite{ rhine2019enhanced}, describes $S=1/2$ antiferromagnets on the square lattice with the spin Hamiltonian $H_{\rm spin}=H_0+H_{B}$, where
\begin{subequations}
\begin{alignat}{2}
\label{eq:mft_spin}
H^{}_0&=\sum_{i<j}J^{}_{ij}\vec{S}^{}_i\cdot\vec{S}^{}_j+\cdots,\\
\label{eq:mft_spin2}
H^{}_B&=J_\chi\sum_{\triangle}\vec{S}^{}_i\cdot \vec{S}^{}_j\times\vec{S}^{}_k-\sum_i\vec{B}^{}_Z\cdot\vec{S}^{}_i.
\end{alignat} 
\end{subequations}
$H_0$ describes nearest-neighbor spin interactions and other possible exchange terms that are invariant under all spacetime symmetries. $H_B$ describes the coupling of the electrons to an applied magnetic field \cite{sen1995large}. The $J_\chi$ term couples to the scalar spin chirality and is induced by the orbital coupling of the magnetic field to the electrons. The value of $J_\chi$ is proportional to
the small magnetic flux penetrating the square lattice. The second term in $H_B$ is the Zeeman coupling of the magnetic field, with the electron magnetic moment absorbed in the definition of $\vec{B}_Z$. Therefore, the physical magnetic field is included in our model through both an orbital coupling ($J_\chi$) and a Zeeman coupling ($\vec{B}_Z$). While $H_{\rm spin}$ could, in principle, also include Dzyaloshinskii-Moriya exchange terms, we do not consider the effect of spin-orbit interactions here. 

Numerical studies of $H_{\rm spin}$ on the square \cite{nielsen2013} and other lattices \cite{haghshenas2019,gong2017,saadatmand2017,wjhu2016,wietek2017,he2015,bauer2014} have found evidence of a chiral spin liquid phase at small nonzero $J_\chi$, and it was argued in Ref.~\onlinecite{rhine2019enhanced} that near a critical spin liquid, $J_\chi$  would be a relevant perturbation leading to semion topological order. Consequently, one finds an enhanced thermal Hall conductivity $\kappa_{xy}$ even in the antiferromagnetic N\'eel state \cite{rhine2019enhanced} stemming from the discontinuity of the zero-temperature thermal Hall response $|\Delta\kappa_{xy}/T|=(\pi/6)(k_B^2/\hbar)$ between the trivial and topological phases \cite{capelli2002}. 
On the other hand, we will find the phonon Hall viscosity to be continuous (but nonanalytic) across this QPT.

\begin{figure}[tb]
\includegraphics[width=\linewidth]{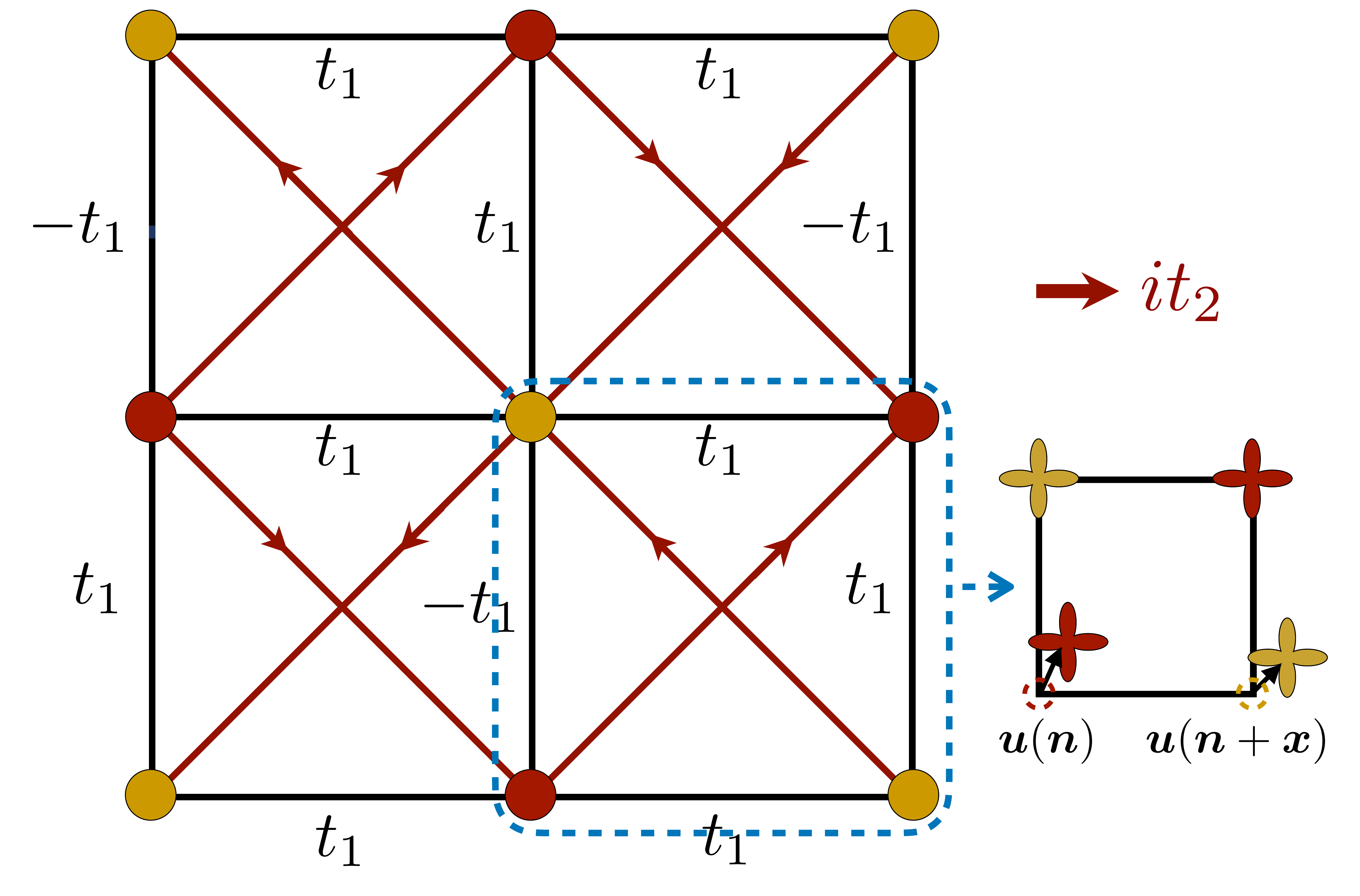}
\caption{The mean-field spinon ansatz defined by Eq.~\eqref{eq:spinon_lattice}, with nearest ($t_1$, black) and second-nearest neighbor ($t_2$, red) hopping matrix elements. The applied magnetic field induces an orbital coupling $it_2$, and there is a uniform $\pi/2$ flux through each elementary triangle. The inset in the bottom-right corner illustrates the bottommost red and yellow atoms with the $d_{x^2-y^2}$ orbitals deviating from their equilibrium positions by $\vec{u}(\vec{n})$ and $\vec{u}(\vec{n}+\vec{x})$, respectively. The result of this deviation can be captured by changing the bond length between the two atoms from the equilibrium length $\vec{r}_0$ to the new length $\vec{r}_0 + (\vec{u}(\vec{n}+\vec{x})-\vec{u}(\vec{n}))$, as discussed further in Sec.~\ref{sec:Lattice_bondstretch}. }
\label{fig:ansatz}
\end{figure}

\subsection{Mean-field theory}
We begin our mean-field analysis by considering the square-lattice N\'eel state as the confining phase of an SU(2) gauge theory of fluctuations
about a $\pi$-flux mean-field state  \cite{wangsenthil2017}. Transforming to the parton representation \cite{affleck88,affleck89}, the spin operator at each site is decomposed as 
\begin{equation}
\label{eq:parton}
\vec{\mathcal{S}}^{}_i=\dfrac{1}{2}f_i^\dagger\vec{\sigma}f^{}_i. 
\end{equation} 
Here, $f_i \equiv (f_{i\uparrow},f_{i\downarrow})^T$ represents the two-component fermionic spinon operator while $\vec{\sigma}$ denotes the Pauli matrices. The mapping from the spin-1/2 Hilbert space to the fermionic one expands the Hilbert space, and we must impose a single-site occupancy constraint in order to remain within the physical Hilbert space. Therefore, the fermionic band structure of spinons is always constrained to be at half-filling. Furthermore, Eq.~(\ref{eq:parton}) has an SU(2) gauge redundancy \cite{lee2006rmp,hermele2007} and a full treatment of $H_{\rm spin}$ would also require analysis of the SU(2) gauge field associated with $f$ \cite{affleckanderson88,coleman88,coleman89}. 

In our mean-field treatment, we ignore the SU(2) gauge fluctuations. Instead, we will be interested in a mean-field saddle point which breaks this SU(2) gauge symmetry down to U(1) \cite{wen2002,wen2004book}. Inserting the parton representation of $\vec{S}_i$ into $H_{\rm spin}$ and mean-field factorizing while respecting the spacetime and gauge symmetries, we obtain the spinon Hamiltonian \cite{wangsenthil2017,wen89,wen2002,scheurer_sachdev_2018,rhine2019enhanced}
\begin{align}
    {H}^{}_{\rm t.b.}=&-\sum_{i<j}\left(t^{}_{ij}f_j^\dagger f^{}_i+t_{ij}^*f_i^\dagger f^{}_j\right)\nonumber\\
    &-\dfrac{1}{2}\sum_{i}(\vec{B}^{}_Z+\zeta^{}_i\vec{N})\cdot f_i^\dagger\vec{\sigma} f^{}_i \label{eq:spinon_lattice}.
\end{align}
Our ansatz for the spinon hopping terms $t_{ij}$ is shown in Fig.~\ref{fig:ansatz}. The nearest-neighbor hopping terms, $t_1$, arise from the factorization of the Heisenberg exchange couplings in $H_0$ [Eq.~(\ref{eq:mft_spin})]. The second nearest-neighbor hopping terms, $\pm it_2$, originate from the scalar spin chirality $J_\chi$ in Eq.~(\ref{eq:mft_spin2}), and they have the same symmetry as the orbital coupling of the electrons to an applied magnetic field orthogonal to the lattice plane. In particular, the field-induced couplings $t_2$ break time-reversal and reflection symmetries but preserve their composition. We have also assumed a nonzero N\'eel order $\vec{N}$, with $\zeta_i =\pm 1$ on the two checkerboard sublattices ($A/B$) of the square lattice. The N\'eel order is temperature-dependent in general but for simplicity, here, we regard $N$ as fixed.  In order to minimize the energy of the antiferromagnet with a Zeeman coupling, we take $\vec{B}_Z\cdot\vec{N}=0$. The Zeeman coupling along the $\hat{\vec{z}}$-axis originates from the perpendicular applied external field, so $\vec{N}$ lies in the $xy$-plane. As there is no spin-orbit coupling, we will perform a rotation in spin space for convenience, so that $\vec{B}_Z \propto \hat{\vec{x}}$ and $\vec{N} \propto \hat{\vec{z}}$. Equation~\eqref{eq:spinon_lattice} can be written in momentum space, with $f_{\vec{k}} \equiv  \sum_i e^{i \vec{k} \cdot \vec{r}_i} f_i / \sqrt{L} = \qty(f_{\vec{k} A \uparrow}, f_{\vec{k} B \uparrow}, f_{\vec{k} A \downarrow}, f_{\vec{k} B \downarrow})^T$, as 
\begin{align}
    H^{}_{\rm t.b.}&=-\sum_{\vec{k}} f_{\vec{k}}^\dagger h^{}_{\vec{k}} f^{\phantom{\dagger}}_{\vec{k}}, \\
    h^{}_{\vec{k}} & = 2 t^{}_1 \cos(k^{}_x)\tau^x-2 t^{}_1 \sin(k^{}_y)\tau^y \nonumber \\
    & + 4 t^{}_2 \sin(k^{}_x) \cos(k^{}_y) \tau^z + \frac{N}{2} \sigma^z \tau^z + \frac{\abs{\vec{B}^{}_Z}}{2} \sigma^x,
    \label{eq:ham_momentum}
\end{align}
where the Pauli matrices acting in sublattice and spin spaces are denoted by $(\tau^x, \tau^y, \tau^z)$ and $(\sigma^x, \sigma^y, \sigma^z)$, respectively.

\begin{figure}[tb]
\includegraphics[width=\linewidth]{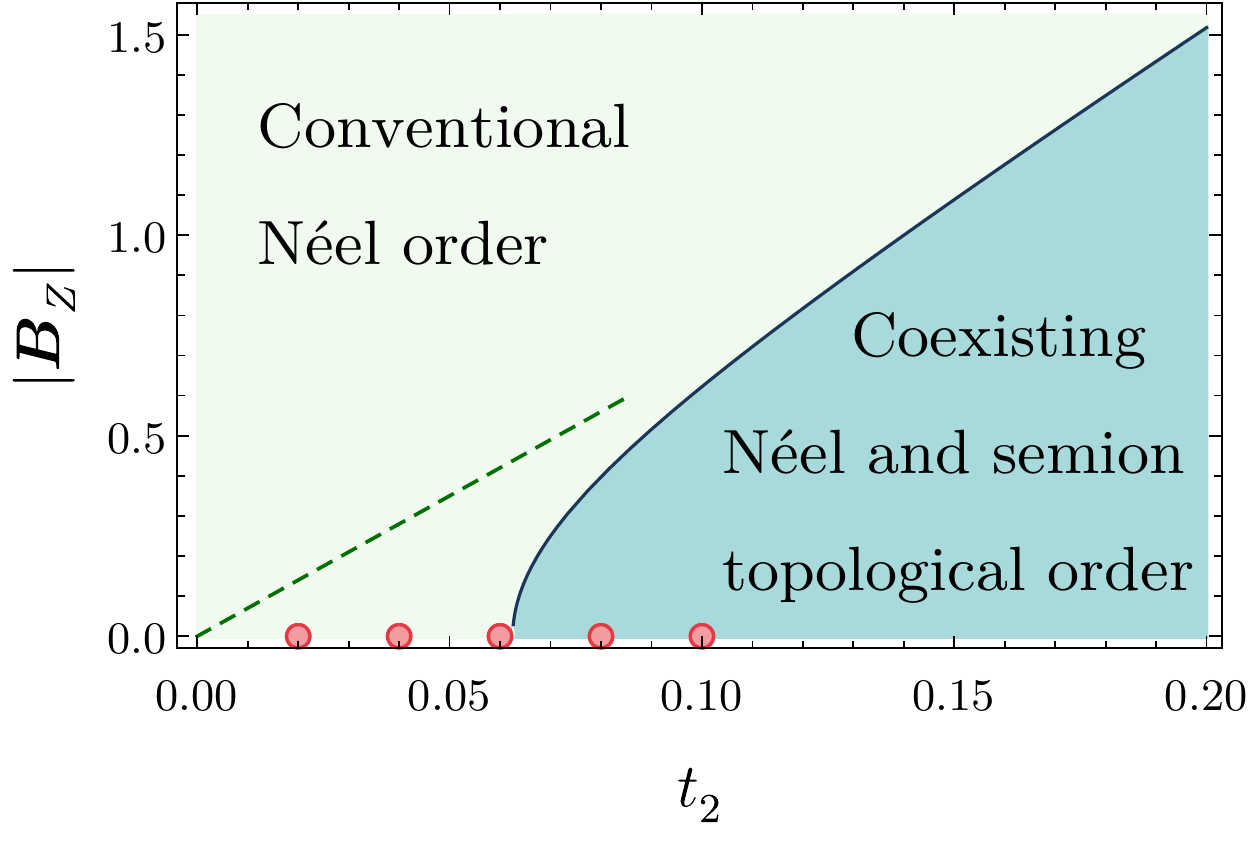}
\caption{\label{fig:PD} The two phases of the spinon mean-field Hamiltonian ${H}_{\rm t.b}$ in Eq.~(\ref{eq:spinon_lattice}) are shown as a function of the second-nearest-neighbor spinon hopping $t_2$ and the strength of the Zeeman field $|\vec{B}_Z|$. Here, we take the N\'eel order to be $\vec{N}=0.5\hat{\vec{z}}$ and measure all energies in units of the nearest-neighbor spinon hopping $t_1$. As discussed in the main text, both $t_2$ and $|\vec{B}_Z|$ are assumed to be linear functions of the applied magnetic field. The red dots show the points for which we plot the temperature dependence of $\eta^H$ in Fig.~\ref{fig:Lattice_fig1}(b), and the dashed green line illustrates the trajectory for which we plot the field dependence of $\eta^H$ in Fig.~\ref{fig:Lattice_fig1}(c).}
\end{figure}

The mean-field phase diagram for this ansatz is sketched in Fig.~\ref{fig:PD}. With our choice of a two-site unit cell, we obtain a total of four spinon bands, which are half filled. When the net Chern number of the occupied bands is zero, one obtains a conventional N\'eel state, and the theory for the gauge fluctuations will have no Chern-Simons term, leading to confinement. However, when the net Chern number of the filled bands is two, we obtain a state with semion topological order coexisting with the N\'eel order. With a fixed N\'eel order, one can thus move between the two phases by tuning the orbital ($t_2$) and Zeeman ($\vec{B}_Z$) couplings of an applied field. We discuss this point further in Sec.~\ref{sec:Lattice_CP}.

While this specific ansatz may appear to break lattice symmetries at first sight, the representation Eq.~(\ref{eq:parton}) is invariant under the local gauge transformations 
\begin{equation}
    f^{}_i\rightarrow f^{}_ie^{i\vartheta_i},\quad t^{}_{ij}\rightarrow t^{}_{ij} e^{i( \vartheta_i-\vartheta_j)}.
\end{equation} 
Accordingly, the representation of lattice symmetries can be supplemented by an appropriate gauge transformation, and so the spinons $f$ form a projective representation of the lattice symmetry group, called the projective symmetry group \cite{wen2002,wen2004book,messio2016}. With this gauge freedom in mind, the spinon lattice model ${H}_{\rm t.b.}$ in \equref{eq:spinon_lattice} indeed preserves all the symmetries of the original spin Hamiltonian $H_{\rm spin}$, as shown in Ref.~\onlinecite{rhine2019enhanced}. We will also consider the projective symmetries in detail when we analyze the continuum spinon theory in Sec.~\ref{sec:continuum}.

In what follows, we consider the total spinon-phonon action
\begin{equation}
\label{eq:total_ham}
     S^{}_{\rm total} = S^{}_{\rm sp}+ S^{}_{\rm ph}+ S^{}_{\rm sp-ph} 
\end{equation} 
in both the lattice and continuum settings. The first term in Eq.~(\ref{eq:total_ham}) is obtained from our ansatz ${H}_{\rm t.b.}$ in Eq.~(\ref{eq:spinon_lattice}), while the second term is obtained from the quadratic phonon action in Eq.~(\ref{eq:quadratic_phonon_action}). On the lattice (Sec.~\ref{sec:Lattice}) we deduce the necessary elastic coupling to the fermions, $S _{\rm sp-ph}$, from geometric bond stretching whereas in the continuum (Sec.~\ref{sec:continuum}), we will derive the allowed elastic coupling to the fermions from symmetry considerations. Our goal will be to integrate out the spinon degrees of freedom to obtain an effective theory for the acoustic phonons (see also Fig.~\ref{fig:feynman_diagram}).

\section{Hall viscosity from spinon couplings to lattice strain fields}\label{sec:Lattice}

Given our tight-binding ansatz in Sec.~\ref{sec:ansatz}, we will model the spinon-phonon coupling through the microscopic deformation of the hopping amplitudes as the result of lattice strain, i.e, bond stretching. As mentioned previously, we assume that the spinons only couple to low-energy phonons with frequencies well below the spinon energy gap. There are two equivalent ways of computing the resulting response to the lattice distortion by bond stretching. The first is to compute the one-loop phonon effective action by integrating out the spinons [shown in Fig.~\ref{fig:feynman_diagram}(b)]; the second is to use the linear response formalism and compute the adiabatic Berry curvature as the result of the variation of the strain field \cite{barkeshli2012,shapourian2015viscoelastic}. We use the first approach here as it more closely makes contact with our later continuum calculations.

\subsection{Geometric coupling through bond stretching}\label{sec:Lattice_bondstretch}
To introduce the method of geometric bond stretching, we will consider a generic tight-binding model
\begin{align}
    {H}^{}_{\rm t.b.} = \sum_{ij} t_{ij}c^\dagger_j  c^{}_i, 
\end{align}
where the hopping amplitude $t_{ij}$ represents the overlap integral between the orbitals at site $i$ and site $j$ with bond length (spatial separation) $\abs{\vec{r}_0}$. In models with multiple orbitals with nontrivial symmetry properties, the change of hopping amplitudes can also have an angular dependence. In our case here, however, Eq.~\eqref{eq:spinon_lattice} is a model with only one type of orbital ($d_{x^2-y^2}$) on each site of the square lattice, so, to leading order, the hopping amplitude only depends on the distance between the two sites. 
Following the approach of Ref.~\onlinecite{shapourian2015viscoelastic}, suppose now that the bond length becomes a variable $\vec{r}$ so that we can introduce a bond stretching of $\delta \vec{r} = \vec{r}- \vec{r}_0$, illustrated in the inset of Fig.~\ref{fig:ansatz}. Assuming that $t_{ij}$ is a smooth function of such small deformations, the hopping amplitude then becomes
\begin{align}
    t(\vec{r}) \equiv t^{}_{\vec{r}_i, \vec{r}_j} \approx t(\vec{r}_0) +  \delta \vec{r}\cdot\eval{\grad t(\vec{r})}_{\vec{r}_0} + \order{\delta \vec{r} ^2}. 
\end{align}
For example, the nearest-neighbor (horizontal) hopping amplitude from site $\vec{n}$ to $\vec{n}+\vec{x}$ is $t_{\vec{n},\vec{n}+\vec{x}}=t(\vec{x})$ where $\vec{x} = a \hat{\vec{x}}$. Letting $u_x(\vec{n})$ and $u_x(\vec{n}+\vec{x})$ be the deformations along the $\hat{\vec{x}}$-axis of the two sites, the hopping amplitude is approximated as
\begin{align}
    t^{}_{\vec{n},\vec{n}+\vec{x}} & \approx t(\vec{x}) + \eval{\pdv{t}{r} }_{\vec{x}}   \big(u^{}_x(\vec{n}+\vec{x})-u^{}_x(\vec{n})\big)+ \order{\delta \vec{r} ^2},\nonumber \\
    & = t(\vec{x}) + a \eval{\pdv{t}{r} }_{\vec{x}} \big( \partial^{}_x u^{}_x\big), \label{eq:bond_stretch_t}
\end{align}
where, on the second line, we have assumed that the lattice distortion is a smooth function on the lattice scale. This is consistent with our assumption of considering only adiabatic spinon-phonon interactions. Carrying through this procedure with our mean-field ansatz defined in Eq.~(\ref{eq:spinon_lattice}), we obtain the modified hopping amplitudes as 
\begin{subequations}
\begin{alignat}{2}
    |t^{}_{\vec{n},\vec{n}\pm \vec{x}}| &\approx t^{}_1   + \lambda^{}_1 \epsilon^{}_{xx} , \label{eq:Lattice_bondStretching} \\
|t^{}_{\vec{n},\vec{n}\pm \vec{y}}| &\approx t^{}_1 + \lambda^{}_1 \epsilon^{}_{yy}, \\
|t^{}_{\vec{n},\vec{n}\pm(\vec{x}+\vec{y})}| &\approx t^{}_2 + \lambda^{}_2 ( \epsilon^{}_{xx} + \epsilon^{}_{yy} + 2 \epsilon^{}_{xy}) , \\
|t^{}_{\vec{n},\vec{n}\pm(\vec{x}-\vec{y})}| &\approx t^{}_2 + \lambda^{}_2 ( \epsilon^{}_{xx} + \epsilon^{}_{yy} - 2 \epsilon^{}_{xy}) \,,
\end{alignat}
\label{AllBonds}\end{subequations}
expressed in terms of the strain tensor $\epsilon_{ij}$ in \equref{DefinitionOfTensors}. The coupling constants $\lambda_j$ are formally given by $\lambda_1 \equiv a\, (dt_1/dr)\vert_{\vec{a}}$ and $\lambda_2 \equiv ({a}/{\sqrt{2}) }(dt_2/dr)\vert_{\sqrt{2}\vec{a}}$ in the bond-stretching picture. Since $\lambda_j$ has the same symmetry as $t_j$, we will take the two to be linearly related; their dimensionless ratio, $\lambda_j/t_j$ will be treated as an unknown, phenomenological parameter.

\begin{figure}[t]
\includegraphics[width=\linewidth]{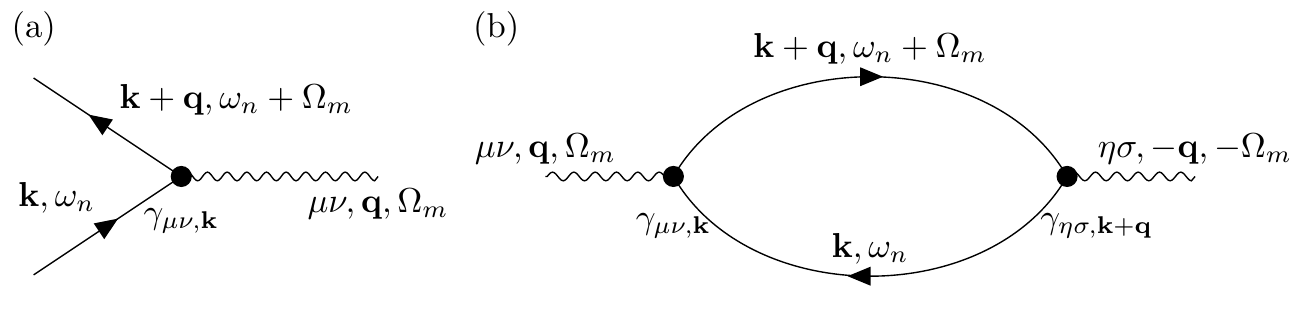}
\caption{(a) Spinon-phonon interaction vertex, as defined by Eq.~\eqref{eq:Lattice_couplingVertices}. (b) The Feynman diagram representing the phonon self-energy, which contributes to the one-loop effective action and determines the phonon Hall Viscosity. Note that we work in the $\vec{q}$\,$=$\,$0$ limit when computing the Hall viscosity on the lattice. For our calculations in the continuum, we also evaluate the same diagram though the precise notations differ.}
\label{fig:feynman_diagram}
\end{figure}

Replacing the fixed hopping amplitudes in \equref{eq:spinon_lattice} by their strain-dependent generalizations in \equref{AllBonds}, one can systematically derive all spinon-phonon coupling terms; for example, the term in Eq.~\eqref{eq:Lattice_bondStretching} leads to a coupling term of the schematic form $f^\dagger \qty( \lambda_1  \epsilon_{xx} ) \qty(\cos (k_x) \tau^x) f$, where the $\cos (k_x)\tau^x$ piece in sublattice space originates from Eq.~\eqref{eq:ham_momentum}. 

\begin{figure}[t!]
\includegraphics[width=\linewidth]{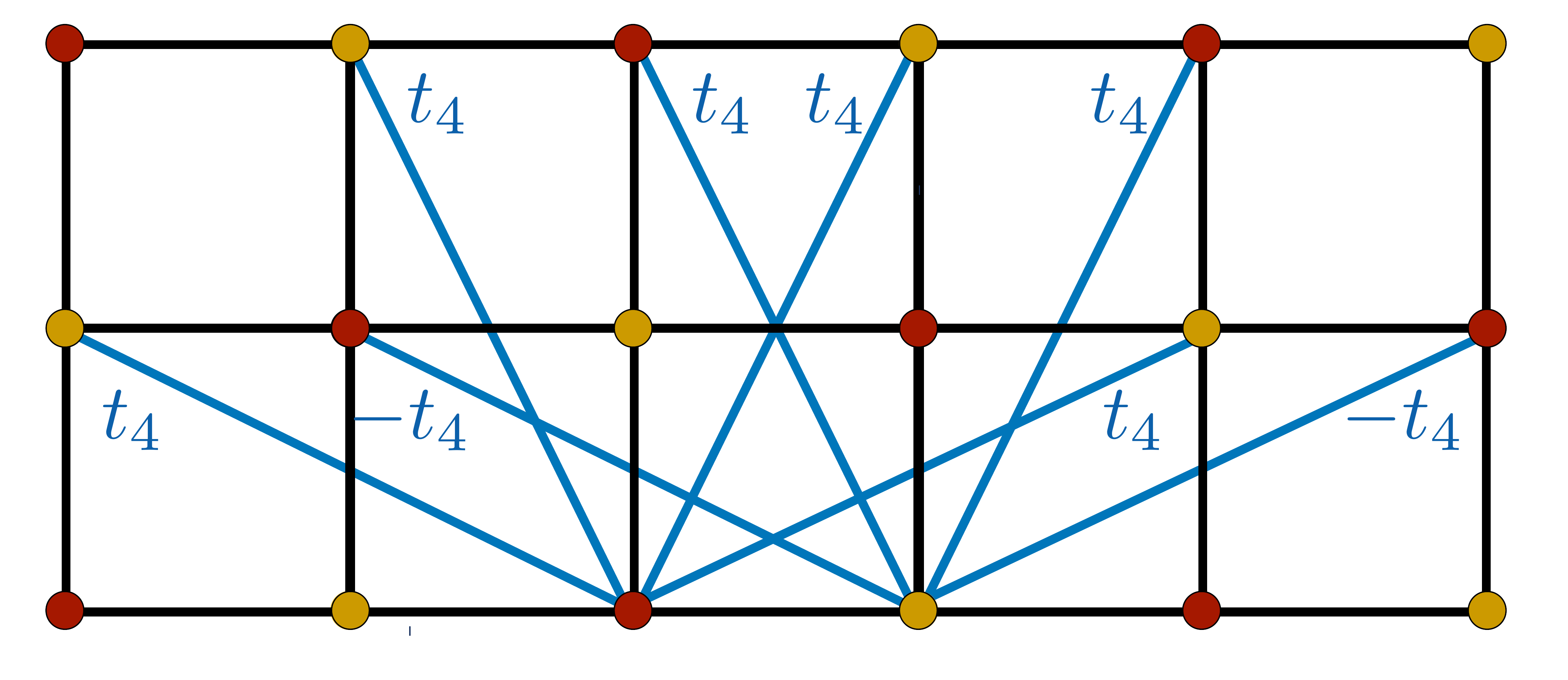}
\caption{Spinon ansatz with fourth-nearest-neighbour (4NN) hopping amplitude $t_4$ allowed by projective symmetry.}
\label{fig:ansatz4nn}
\end{figure}

Before listing the precise structures of all of these couplings, we comment on further-neighbor couplings. While we only include terms involving up to second nearest-neighbors (2NN) in our spinon ansatz, higher-neighbor terms are still allowed by symmetry. Usually, these couplings are not necessary as they are expected to be weak in magnitude and can often effectively be taken into account by renormalizing the NN or 2NN terms. However, it turns out  that additional fourth-nearest-neighbour (4NN) terms are crucial for our analysis: while their coupling strengths may be numerically small, their induced phonon coupling alters the divergent behavior of $\eta^H$ at the critical point, as we will see below and also consistently reproduce later in the continuum analysis of \secref{sec:continuum}.
To include their effects, we use the projective symmetry of Eq.~\eqref{eq:spinon_lattice} to find allowed 4NN terms with hopping strength $t_4$, as shown in Fig.~\ref{fig:ansatz4nn}. Following the bond-stretching procedure, we define an analogous parameter $\lambda_4 \equiv ({a}/{\sqrt{5}) }(dt_4/dr)\vert_{\sqrt{5}\vec{a}}$. As the $t_4$ coupling will not modify the critical behavior of the spinon Hamiltonian, we will take $t_4\rightarrow 0$ so that it only enters through the spinon-phonon coupling Hamiltonian.

Moreover, while it is formally possible to consider time-dependent deformations $\dot{\vec{u}}$, we will not include them in our analysis since these terms will be suppressed by the ratio of the sound velocity to the Fermi velocity ($\sim t_1$). We comment on potential interesting effects from these terms in Appendix~\ref{app:EtaM_derivation}. Lastly, we note that couplings similar to the ones induced by $\lambda_4$ can also arise from bond stretching in a multiorbital model; for coupling between $s-p$ and $d-p$ orbitals, Eq.~\eqref{eq:bond_stretch_t} would include terms that take into account the relative rotation between sites. 

Summarizing all the relevant bond stretching coupling terms, we can write the spinon-phonon coupling as $K_{\vec{k},\Omega} = \gamma_{\mu\nu, \vec{k}} \epsilon_{\mu\nu}(\Omega)$, which couples the spinon operators: 
\begin{align}\label{eq:Lattice_couplingVertices}
     S^{}_{\rm{sp-ph}}&=  \frac{1}{L^2 \beta^2} \sum_{ \omega, \Omega, \vec{k}}f_{\vec{k}, \omega+\Omega}^\dagger K^{}_{\vec{k},\Omega} f^{}_{\vec{k}, \omega}.
\end{align}
This interaction vertex is displayed in Fig.~\ref{fig:feynman_diagram}(a), but we will take the limit in which the strain field, $\epsilon$, carries no momentum, as terms dependent on the phonon momentum will lead to higher order, anharmonic viscosity terms in the phonon effective action. The interaction vertex $\gamma_{\mu\nu, \vec{k}}$ then reads as \begin{widetext}
    \begin{align}
    \gamma^{}_{\mu\nu, \vec{k}} \equiv \sum_i \gamma_{\mu\nu,\vec{k}}^i \tau^i= \begin{cases}
      - 2  \lambda^{}_1 \cos (k^{}_x) \tau^x - 4  \lambda^{}_2 \sin(k^{}_x)\cos(k^{}_y) \tau^z , &  \mu\nu = xx \\
      - 2  \lambda^{}_1 \sin (k^{}_y) \tau^y - 4  \lambda^{}_2 \sin(k^{}_x)\cos(k^{}_y) \tau^z , &  \mu\nu = yy \\
      - 8  \lambda^{}_2 \cos(k^{}_x)\sin(k^{}_y) \tau^z  -16  \lambda^{}_4 \qty[\cos(k^{}_y) \sin(2 k^{}_x) \tau_y +\sin(2 k^{}_y) \sin(k^{}_x) \tau_x],  & \mu\nu = xy 
    \end{cases} \label{eq: Lattice_bondStretching_couplingVertex}
\end{align} 
\end{widetext}
where $\gamma_{\mu\nu,\vec{k}}^i$ defines the coefficient multiplying the Pauli matrix $\tau^i$ in $\gamma_{\mu\nu, \vec{k}}$.

\subsection{Evaluation of the phonon self-energy}\label{sec:lattice_phonon_bubble}
Given the couplings from the previous section, we will now integrate out the spinons ($\psi$) from the total partition function \cite{barkeshli2012, serbynlee2013, yeperkins2020},
\begin{align}
    \mathcal{Z} &= \int \mathcal{D}\bar{\psi}\, \mathcal{D}\psi\, \mathcal{D}u\: e^{- \qty(  S _{\rm ph}(u) +  S _{\rm sp}(\bar{\psi},\psi) +  S _{\rm sp-ph}(\bar{\psi},\psi, u) )},  \nonumber \\
    & = \int \mathcal{D}u\;e^{- S _\text{eff}(u) }. \label{eq:effective_action}
\end{align} 
This is equivalent to evaluating the phonon self-energy to lowest-order in the interaction couplings and leads to a term in the phonon effective action:
\begin{align} \delta S^{}_{\rm{eff}}
    = & - \frac{1}{2 L^2 \beta^2} \sum_{\substack{ \omega_n, \Omega_m, \vec{k}}} \Tr \left[K^{}_{\vec{k},i \Omega_m} G (\vec{k},i \omega^{}_n) \right.\nonumber \\
& \times \left.K^{}_{\vec{k},-i \Omega_m} G (\vec{k},i\omega^{}_n+ i\Omega^{}_m ) \right]\label{eq:Lattice_EffAction}
 \end{align}
which can be represented by the Feynman diagram in Fig.~\ref{fig:feynman_diagram}(b) with the external momenta $\vec{q}=0$. In the absence of the Zeeman field ($\vec{B}_Z = 0$), the two spin sectors are decoupled so we write the block-diagonal spinon Green's function for the spin up ($+$) and spin down ($-$) sectors as
\begin{align}
    G^{}_{\pm} (\vec{k},i\omega^{}_n) = \dfrac{i\omega^{}_m \mathbb{I}+\mathbf{H}_{\pm,\vec{k}} \cdot\boldsymbol{\tau}}{(i\omega^{}_m)^2-\mathbf{H}_{\pm,\vec{k}}^2},
\end{align}
in which $\mathbf{H}$ is defined from the momentum space Hamiltonian in Eq.~\eqref{eq:ham_momentum},
\begin{align}
    \mathbf{H}^{}_{\pm,\vec{k}}\equiv (- 2 t^{}_1 \cos{k^{}_x},  2 t^{}_1 \sin{k^{}_y} ,- 4 t^{}_2 \sin{k^{}_x}\cos{k_y} \mp \frac{N}{2} ). 
\end{align} 
Leaving the details of the derivation to Appendix~\ref{app:EtaH_derivation}, Eq.~\eqref{eq:Lattice_EffAction} leads to a Hall viscosity 
\begin{align}
    &\eta^H = -\frac{1}{L^2} \sum_{\vec{k}, \pm}  \left( \frac{ \qty(1-2  n_F(\abs{\mathbf{H}_\pm})+2\abs{\mathbf{H}_\pm} n^\prime_F(\abs{\mathbf{H}_\pm})) 
   }{4  \abs{\mathbf{H}_\pm}^3}   \right) \nonumber \\
   & \times\qty[ \mathbf{H}_\pm^y \gamma_{xx}^z \gamma_{xy}^x + \mathbf{H}_\pm^z  \gamma_{xx}^x \gamma_{xy}^y-  \mathbf{H}_\pm^x  \gamma_{xx}^z \gamma_{xy}^y -  \mathbf{H}_\pm^y  \gamma_{xx}^x \gamma_{xy}^z], \label{eq:Lattice_EtaH}
\end{align}
where $n_F(E) = 1/(1+ e^{E/T})$ denotes the Fermi distribution function with chemical potential at $0$, and $n^\prime_F(E)$ denotes its first derivative with respect to $E$. We have also suppressed the momentum indices of $\mathbf{H}_{\pm,\vec{k}}$ and $\gamma_{\mu\nu,\vec{k}}^i$ for ease of notation. The terms multiplying the thermal factor in Eq.~\eqref{eq:Lattice_EtaH} should be thought of as an effective Berry curvature for the phonon Hall viscosity, with the summation being over occupied spinon states.

\begin{figure*}[h]
\includegraphics[width=\linewidth]{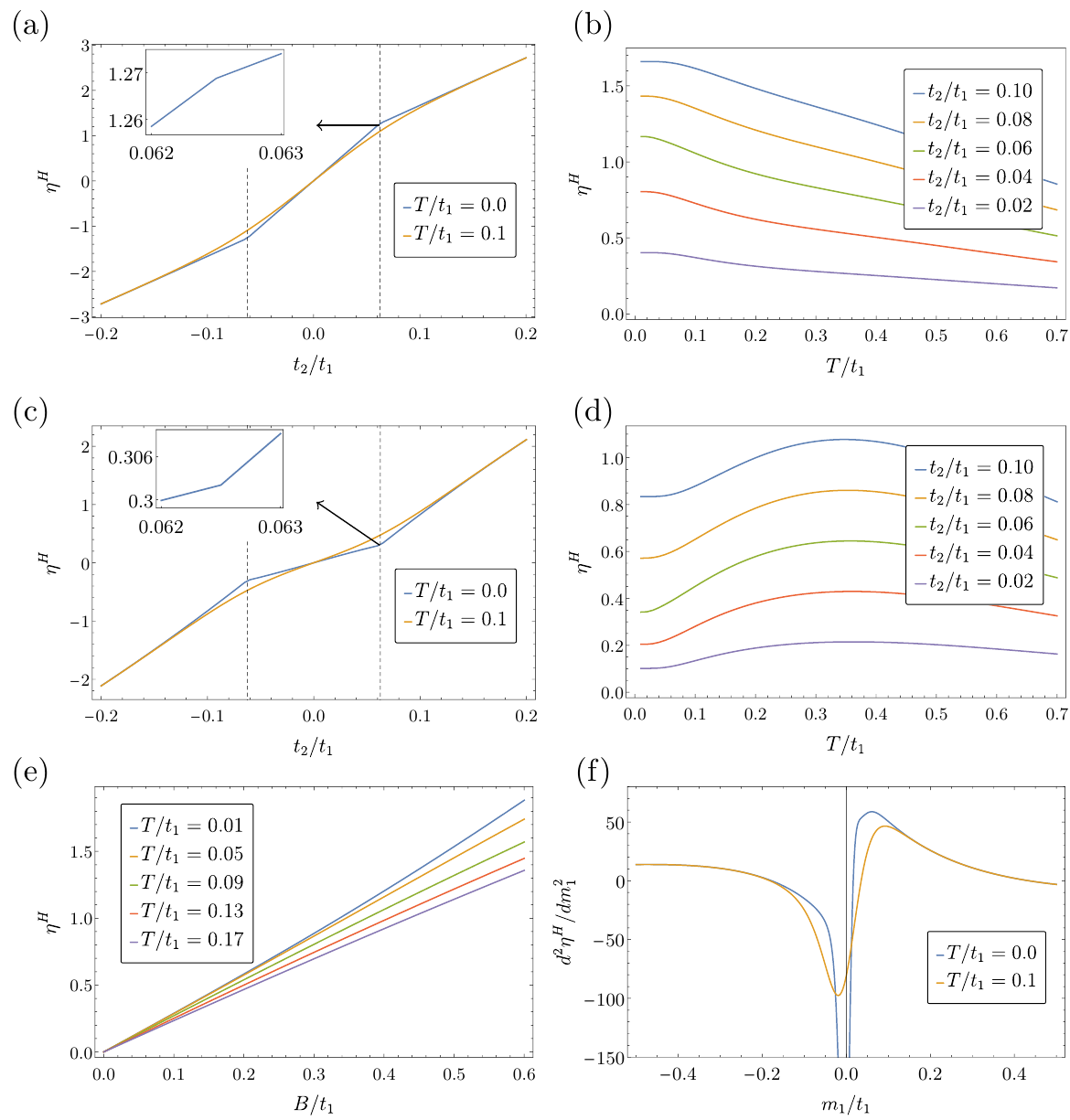}
\caption{Hall viscosity as functions of $t_2, T, B$ and its derivative with respect to $m_1$. We set $N = 0.5$, $\lambda_1 = t_1$ and $\lambda_2 = 0.5 t_2$. In (a,b) and (e,f), we choose $\lambda_4/t_1 = -0.1$. In (c,d), we choose $\lambda_4/t_1 = 0.1$ for comparison. (a) Hall viscosity as a function of $t_2$ for temperatures $T/t_1 = 0, 0.1$ and $\vec{B}_Z=0$. The dashed lines indicate critical points at $N = \pm 8 t_2$. The inset shows the kink at $T=0$ after zooming in, signaling a QPT as discussed in the main text. (b) Hall viscosity $\eta^H$ as a function of temperature for different orbital coupling $t_2$ and $\vec{B}_Z=0$, shown by the red dots in phase diagram Fig.~\ref{fig:PD}. (c, d) The same as in (a) and (b), respectively, but with the opposite sign of $\lambda_4$. (e) Field dependence of $\eta^H$ for different $T$. As discussed in the main text, the orbital and Zeeman couplings scale linearly with the applied external field, and here, we take $B =|\vec{B}_Z| = 7 t_2$, which is shown by the dashed trajectory in phase diagram Fig.~\ref{fig:PD}. (f) The divergence in the second derivative of $\eta^H$ with respect to mass $m_1$ near the critical point. The blue curve shows that the mass derivative of the Hall viscosity $d^2 \eta^H / d{m_1^2}$ evaluated at $m_1 = 0$ diverges at $T=0$; the yellow curve demonstrates that $d^2 \eta^H / d{m_1^2}$ has no true divergence at finite $T$. }
\label{fig:Lattice_fig1}
\end{figure*}

The phonon Hall viscosity is shown in Figs.~\ref{fig:Lattice_fig1}(a)--(e). Let us first concentrate on the zero Zeeman field limit $\vec{B}_Z = 0$ described by Eq.~\eqref{eq:Lattice_EtaH}. Figures~\ref{fig:Lattice_fig1}(a) and \ref{fig:Lattice_fig1}(b) show the Hall viscosity for $\lambda_4$\,$=$\,$-0.1$ while Figs.~\ref{fig:Lattice_fig1}(c) and \ref{fig:Lattice_fig1}(d) display the viscosity for $\lambda_4$\,$=$\,$0.1$. Recall, as mentioned previously, that we take $\lambda_1\propto t_1$ and $\lambda_2\propto t_2$. We first observe that $\eta^H$ is an odd function of $t_2$: this property arises from the second line of Eq.~\eqref{eq:Lattice_EtaH} via either the Green's function component $\mathbf{H}^z$ or the interaction vertex $\gamma_{xy}^i$. The viscosity vanishes when $t_2 = 0$ in consistency with the fact that it can only be nonzero when time-reversal and mirror symmetries are broken. The viscosity also monotonically increases with increasing $t_2$ across the critical points $N = \pm 8 t_2$. As discussed in Sec.~\ref{sec:ansatz}, $t_2$ originates from the orbital coupling of the magnetic field, so tuning $t_2$ should be understood as tuning the magnetic flux threading the square lattice. Furthermore, from Figs.~\ref{fig:Lattice_fig1}(a) and \ref{fig:Lattice_fig1}(c), we notice that although the viscosity is continuous, it exhibits a kink at zero temperature at the quantum critical point, signaling a discontinuous first derivative. The exact difference in the slope of $\eta^H$ on either side of the critical point is nonuniversal and depends on the choice of couplings. In our ansatz, we see that a negative (positive) $\lambda_4$ leads to a smaller (larger) slope for $\eta^H$ in the topological phase.

The behavior of $\eta^H$ as a function of temperature is also of experimental relevance. Figures~\ref{fig:Lattice_fig1}(b) and \ref{fig:Lattice_fig1}(d) illustrate the temperature dependence of $\eta^H$ for different values of $t_2$ (while keeping $\vec{B}_Z=0$), which are indicated by the red dots in the phase diagram of Fig.~\ref{fig:PD}. We observe, in both cases, a plateau of $\eta^H$ at small $T$, which scales with the distance of $t_2$ from the critical point (here, $t_{2, \rm c}= N/8 = 0.0625$). From the plots, the extent of the plateau can be seen to be the smallest for $t_2=0.06$ and increases with changing $t_2$ in either direction away from the critical value. The plateau originates from the spinon energy gap, whose scale is set by $|t_2-t_{2, \rm c}|$. At temperatures below this gap, thermal excitations fail to excite higher spinon bands so we expect $\eta^H$ to retain its zero-temperature behavior. 

An interesting feature of the temperature dependence sketched in Figs.~\ref{fig:Lattice_fig1}(b) and \ref{fig:Lattice_fig1}(d) is that at intermediate temperatures above the energy gap, there is a peak in the viscosity for $\lambda_4$\,$=$\,$0.1$ but not for $\lambda_4$\,$=$\,$-0.1$. This peak is nonuniversal, being dependent on our choice of parameters, but its behavior can actually be understood from the behavior of the kink in $\eta^H$ across the QPT. Intuitively, this can be seen as follows. In passing through the QPT, the effective Berry curvature is exchanged between the highest occupied and lowest unoccupied bands when the spinon gap closes. This is similar in essence to the process of changing temperature, which also involves accessing the effective Berry curvature of the lowest-energy unoccupied spinon bands as $\eta^H$ gains (loses) Berry curvature from the unoccupied (occupied) bands due to thermal excitations. For the case of $\lambda_4=-0.1$, the slope of $\eta^H$ with respect to $t_2$ decreases across the QPT. At the QPT, Berry curvature is exchanged between the occupied and unoccupied bands, so at a fixed $t_2$, a similar redistribution of the Berry curvature between the occupied and unoccupied bands should decrease $\eta^H$. This is exactly what occurs at intermediate temperatures because of thermal excitations, and geometrically, this Berry curvature exchange deforms the viscosity towards the secant through the kink, as illustrated in Fig.~\ref{fig:Lattice_fig1}(a). Therefore, it is expected that $\eta^H$ decreases with increasing temperature. A similar analysis for $\lambda_4=0.1$ with the kink in Fig.~\ref{fig:Lattice_fig1}(c) predicts that $\eta^H$ will increase to a local maximum at intermediate temperatures as thermal excitations capture larger Berry curvature contributions from the lowest unoccupied bands. Regardless of our choice of couplings, however, the Hall viscosity will eventually decay to zero at high temperatures because the Berry curvatures from states at all energies will then contribute, and the net curvature from all spinon bands is necessarily zero.

Under a nonzero Zeeman coupling, we expect the location of the critical points to be renormalized without any qualitative changes in the nature of the QPT \cite{rhine2019enhanced}. In Fig.~\ref{fig:Lattice_fig1}(e), we turn on the Zeeman coupling $\vec{B}_Z$. As discussed below Eq.~\eqref{eq:mft_spin}, $J_\chi$ and $\vec{B}_Z$ are proportional to an externally applied, perpendicular field. Since $t_2$ arises from $J_\chi$, for simplicity, we take $t_2$ and $\vec{B}_Z$ to be linear functions of the applied out-of-plane magnetic field $B$, with $B=|\vec{B}_Z|=7t_2$ as shown by the dashed trajectory in Fig.~\ref{fig:PD}. We find that $\eta^H$ scales linearly with $B$ at small field strengths.

\subsection{Hall viscosity near the spinon critical point}\label{sec:Lattice_CP}
Compared to the quantized thermal Hall conductivity or the ordinary Hall conductance, the Hall viscosity plotted in Fig.~\ref{fig:Lattice_fig1} is continuous and, at first sight, does not seem to encode any signatures of a QPT. However, as seen in Fig.~\ref{fig:Lattice_fig1}(a), it is possible for the derivatives of the Hall viscosity to have a discontinuity or divergence at the critical point.

In the mean-field ansatz given in Eq.~\eqref{eq:spinon_lattice}, by choosing appropriate mean-field orbital coupling parameters $t_2$ and N\'eel order $\vec{N}$\,$=$\,$N \hat{\vec{z}}$, one can tune across the topological phase transition. In particular, at $\vec{B}_Z$\,$=$\,$0$, the critical points at $N$\,$=$\,$\pm 8 t_2$ describe the transition between a confining N\'eel state and a state where the N\'eel order coexists with a chiral spin liquid. At both critical points, the spectra have pairs of Dirac cones at $\pm \mathbf{Q}$ where $\mathbf{Q} = \qty(\pi/2, 0)$. For example, when $N =  8 t_2$, fermions in the spin-down ($-$) sector have a Dirac cone at $\mathbf{Q}$ so that $\vert\mathbf{H}_{-, \mathbf{Q}}\vert = 0$; a similar statement follows for the spin-up ($+$) sector. To examine $\eta^H$ near the QPT, we expand the spinon momentum around $-\mathbf{Q}$ as $\vec{k} = -\mathbf{Q} + \vec{q}$ for small momentum $\vec{q}$. Then, we find---to leading order in $\vec{q}$---for $\eta^H$ at $T=0$:
 \begin{align}
        \lambda^{}_4 \qty[ \frac{\qty(t^{}_2 - \frac{N}{8})q_x^2  \lambda^{}_1 -   t^{}_1 \lambda^{}_2 q^2 }{8\, \abs{q^2+ 4 \qty(t^{}_2-\frac{N}{8})^2}^{3/2}} + \frac{\qty(t^{}_2 + \frac{N}{8})q_x^2  \lambda^{}_1 -   t_1 \lambda^{}_2 q^2}{64\, \abs{t^{}_2 + \frac{N}{8}}^3} ]. \label{eq: CriticalPoint_minusQExpansion} 
 \end{align}
As mentioned earlier, we observe that the nonvanishing leading terms above arise from the 4NN spinon-phonon couplings in Eq.~\eqref{eq: Lattice_bondStretching_couplingVertex}. The second term in Eq.~\eqref{eq: CriticalPoint_minusQExpansion} vanishes as we approach $-\mathbf{Q}$. The first term appears divergent but is actually finite when we take into account the summation over momentum, which comes with measure $\abs{\vec{q}} d\abs{\vec{q}}$. 

While $\eta^{H}$ seems well-behaved, its derivatives with respect to the time-reversal-breaking $t_2$ can have singularities and signal a QPT of the spinons. It is convenient to rewrite our expression as a function of the Dirac masses $m_{1,2} = 2t_2 \mp \frac{N}{4}$, which vanish at the critical points. For instance, taking the second derivative of $\eta^H$ with respect to $m_1$ at the critical point $m_1 = 0$ (and $\vec{k} \approx -\mathbf{Q} + \vec{q}$) leads to a $\delta$-function divergence 
\begin{align}
  \pdv[2]{\eta^H}{m_1} \appropto \lambda^{}_4 m^{}_2 \sum_{\vec{q}} \pdv[2]{}{m_1}\frac{q^2}{ \abs{\mathbf{H}^{}_{+, \vec{k}}}^3}\sim  m^{}_2\partial_{m^{}_1}^2|m^{}_1| \xrightarrow[m^{}_1 \rightarrow 0]{} \infty, \label{eq:Lattice_HallVisc_CrPtDeriv}
\end{align} where we have written $\lambda_2\propto t_2\propto m_1+m_2$. As previously noted, the second derivative's divergence manifests as a kink in $\eta^H$ at the QPT. Note that without a nonzero $\lambda_4$, the singularity in $
\eta^H$ would only show up in its \textit{fourth} derivative. 
The divergent behavior of $\eta^H$ is present only at zero temperature, as illustrated in Fig.~\ref{fig:Lattice_fig1}(f). In the limit of $N\rightarrow 0$, the two Dirac masses coincide ($m_1=m_2$), and $\eta^H$ is better behaved, with the divergence appearing in the third derivative. This is actually the behavior seen in previous works \cite{shapourian2015viscoelastic,hughes2013long} that explored the case of two-orbital Chern insulators on the square lattice; we discuss this point further in Sec.~\ref{sec:continuum_disc}.

\section{Hall viscosity from spinon couplings to continuum strain fields}
\label{sec:continuum}
In previous sections, we studied how phonon chirality could emerge from an underlying chiral spin liquid on the square lattice. The calculation of the continuum phonon Hall viscosity is qualitatively the same as that for the lattice phonon Hall viscosity: we begin by defining the spinon Hamiltonian $H_{\rm sp}$ and symmetry considerations constrain the allowed spinon-phonon couplings. However, instead of taking into account lattice displacements through the strain dependence of tight-binding parameters, we will see how phonon couplings emerge from a projective symmetry analysis of the underlying spin liquid. In particular, the lattice space group can have significant effects on the topological quantization of the phonon Hall viscosity. Our representation-theoretic approach follows Ref.~\onlinecite{serbynlee2013}. We also note that the symmetry-based approach to electron-phonon interactions has been well studied in the case of graphene \cite{rachel2016,basko2009,neto2009rmp,guinea2009,levy2010,manes2007}.

\subsection{Continuum low-energy theory}\label{sec:continuum_low_energy} 
\begin{figure}[tb]
\includegraphics[width=\linewidth]{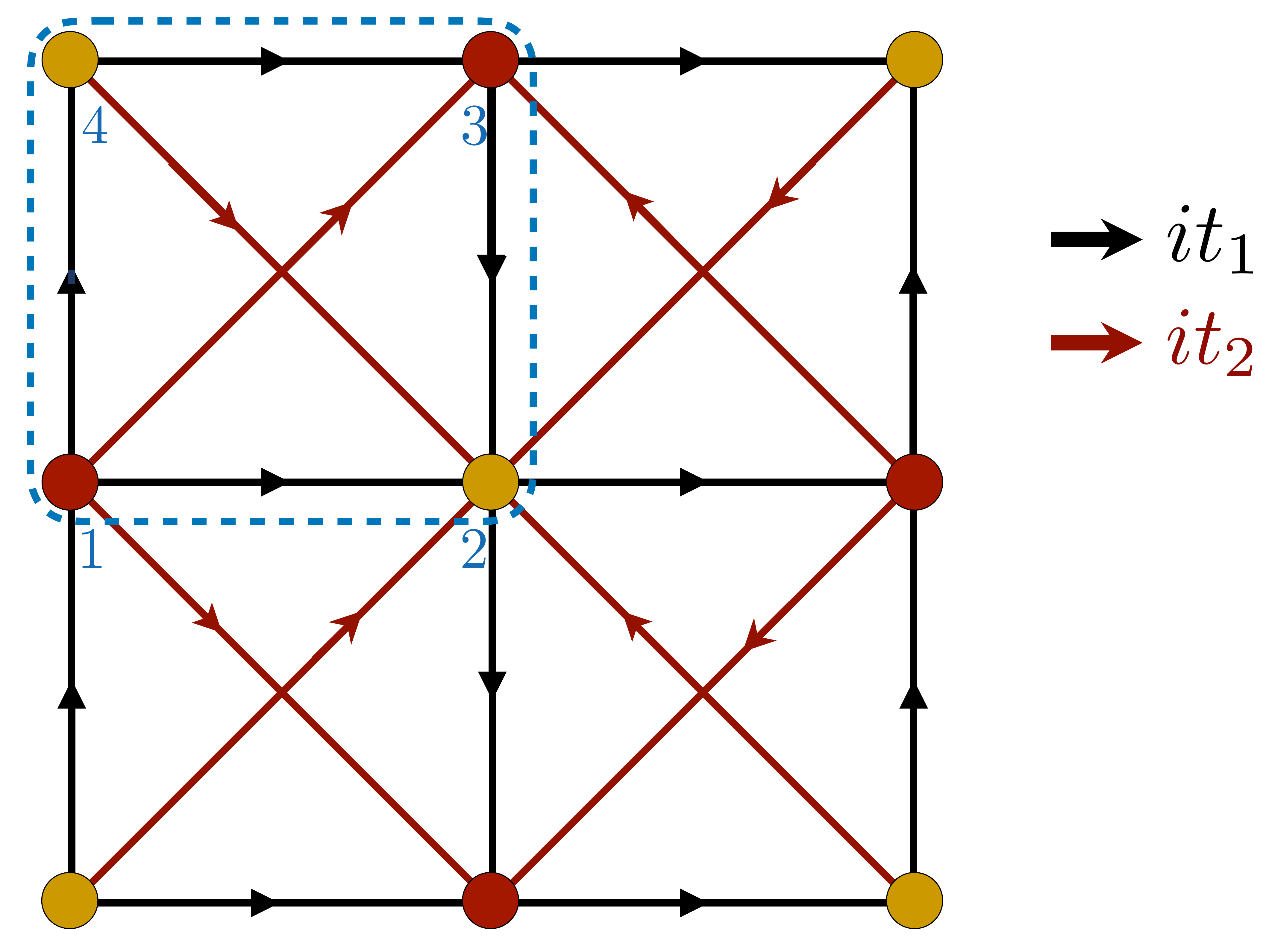}
\caption{The nearest- ($t_1$, black) and second-nearest-neighbor ($t_2$, red) hopping matrix elements for the ansatz in Sec.~\ref{sec:continuum}. It is gauge equivalent to the mean-field ansatz in Fig.~\ref{fig:ansatz} using the transformations outlined in Eqs.~(\ref{eq:gauge1}) and (\ref{eq:gauge2}).}
\label{fig:ansatz_gauge}
\end{figure}

To define our continuum theory, we begin with our original square-lattice N\'eel state $\vec{N}=N\hat{\vec{z}}$, given by $H_{\rm t.b.}$ in Eq.~\eqref{eq:spinon_lattice}. We will also work in the regime of no Zeeman coupling, $\vec{B}_Z=0$.
While the low-energy theory and projective symmetry group of $H_{\rm t.b.}$ was already studied in Ref.~\onlinecite{rhine2019enhanced}, we will find it convenient to first perform a local U(1) gauge transformation in order to match the $\pi$-flux ansatz considered in Ref.~\onlinecite{serbynlee2013}. Our new ansatz, which is nevertheless gauge equivalent to Eq.~\eqref{eq:spinon_lattice}, will have different couplings. The resulting projective symmetries \cite{hermele2004} realized on the low-energy continuum fields will dictate the allowed spinon-phonon interactions. First, we consider a position-dependent gauge transformation of $H_{\rm t.b.}$ in Eq.~\eqref{eq:spinon_lattice}, \begin{subequations}
\begin{alignat}{2}
    f^{}_{\vec{n}}&\rightarrow e^{i\pi n_1/2}f^{}_{\vec{n}}\quad&&\textnormal{for $n^{}_2$ even},\label{eq:gauge1}\\f^{}_{\vec{n}}&\rightarrow e^{i\pi/2}e^{i\pi n_1/2 }f^{}_{\vec{n}}\quad&&\textnormal{for $n^{}_2$ odd},\label{eq:gauge2}
\end{alignat} 
\end{subequations} where $\vec{n}\equiv n_1\vec{x}+n_2\vec{y}$. As a result, our nearest-neighbor spinon hopping terms are given by 
\begin{align}t^{}_{\vec{n},\vec{n}+\vec{x}}=i,\quad t^{}_{\vec{n},\vec{n}+\vec{y}}=(-1)^{n_1}i,\label{eq:continuum_t1}
\end{align}
with second-nearest-neighbor chiral couplings
\begin{subequations}
\begin{alignat}{2}
    t^{}_{\vec{n},\vec{n}\pm \vec{x}\pm \vec{y}}&=it^{}_2\quad&&\textnormal{for $n_1$ even},\label{eq:continuum_t2_1}\\t^{}_{\vec{n},\vec{n}\pm \vec{x}\pm \vec{y}}&=-it^{}_2\quad&&\textnormal{for $n_1$ odd}.\label{eq:continuum_t2_2}
\end{alignat} 
\end{subequations} Now, we relabel our unit cell with four sites as in Fig.~\ref{fig:ansatz_gauge}. The resulting Bravais lattice vector is $\vec{r}=r_1\vec{a}_1+r_2\vec{a}_2$, with $r_1,r_2 \in \mathbb{Z}$ labeling the unit cell, and $\vec{a}_1=2\vec{x}$, $\vec{a}_2=2\vec{y}$. The full form of the Hamiltonian is given in Appendix~\ref{app:psg_continuum}.

Within the Brillouin zone $k_x,k_y\in[-\pi/2,\pi/2)$, our new (but gauge-equivalent) Hamiltonian has degenerate Dirac points at $\Gamma=(0,0)$. Near $\Gamma$, the dispersion can be described by four two-component ($s$\,$=$\,$1,2$) Dirac fermions $\psi^s_{\alpha\sigma}$. The four ``flavors'' $(\alpha,\sigma)$ are associated with the two spin polarizations, $\sigma$\,$=$\,$\uparrow,\downarrow$, and an additional valley index $\alpha$\,$=$\,$1,2$.  
In the following, we will suppress the (sublattice) spinor index $s$ of $\psi^s_{\alpha\sigma}$. 
We can perform an expansion of the momentum-space Hamiltonian [see Eq.~(\ref{eq:continuumH_full})] around $\Gamma$ using the continuum spinor fields $\psi_{\alpha\sigma}$: 
\begin{subequations}\label{eq:continuum_fields}
\begin{alignat}{2}
    \psi^{}_{1\sigma}(\vec{k})&\sim \dfrac{1}{\sqrt{2}}\begin{pmatrix}if^{}_{\vec{k}2\sigma}+f^{}_{\vec{k}4\sigma}\\-if^{}_{\vec{k}1\sigma}-f^{}_{\vec{k}3\sigma}\end{pmatrix},\\
    \psi^{}_{2\sigma}(\vec{k})&\sim \dfrac{1}{\sqrt{2}}\begin{pmatrix}if^{}_{\vec{k}3\sigma}+f^{}_{\vec{k}1\sigma}\\-if^{}_{\vec{k}4\sigma}-f^{}_{\vec{k}2\sigma}\end{pmatrix},
\end{alignat} 
\end{subequations}
from which the resulting Dirac Hamiltonian is 
\begin{align}
&H^{}_{\rm Dirac}=\int\dfrac{d^2\vec{k}}{(2\pi)^2}\psi_{\alpha\sigma}^\dagger\bigg[  v^{}_F \left(k^{}_x\tau^x+k^{}_y\tau^y\right) \label{eq:continuum_diracHam}\\ 
\nonumber & \left. -2
t^{}_2(k^{}_x\mu^x\tau^x+k^{}_y\mu^y\tau^y)
 + 
 2\qty(2t^{}_2 \tau^z - \frac{N}{4} \sigma^z \mu^z \tau^z)\right]\psi^{}_{\alpha\sigma},
\end{align}
where we have labeled $v_F=2t_1$. We have defined the Pauli matrices $\tau$ to act on the spinor (sublattice) indices, $\mu$ to act on the valley indices $\alpha$, and $\sigma$ to act on the spin indices. The continuous fields also realize a projective representation of our lattice symmetries, the details of which are summarized in Appendix~\ref{app:psg_continuum}. Away from the critical points, the Dirac fermions $\psi_{\alpha \sigma}$ are gapped with a mass $m_{1,2}=2t_2 \mp N/4$ given by a combination of the orbital current $t_2$ and the N\'eel order, as in Sec.~\ref{sec:Lattice_CP}. Therefore, when $t_2\approx N/8$ close to the critical point, we can safely integrate out the two higher-energy bands to obtain the effective spinon Hamiltonian
\begin{equation}
\label{eq:lowenergyspinon}
H^{}_{\rm sp}=2\int\dfrac{d^2\vec{k}}{(2\pi)^2}\Psi_a^\dagger\left(k^{}_x\tau^x+k^{}_y\tau^y+m\tau^z\right)\Psi^{}_a,
\end{equation} 
where we have set $t_1=1$ and defined 
\begin{equation}
m\equiv m^{}_1=2t^{}_2-N/4,\quad \Psi^{}_a(k) =
    \begin{cases}
        \psi^{}_{1\uparrow}(\vec{k}) & a=1, \\
        \psi^{}_{2\downarrow}(\vec{k}) & a=2, \\
    \end{cases}
\end{equation}
and, as previously mentioned, the Pauli $\tau$ matrices only act on the spinor indices. From here on, we will also denote the higher-energy Dirac mass as $M\equiv m_2=2t_2+N/4$. Interestingly, the effects of the orbital current $t_2$ and N\'eel order $N$ counteract each other in the low-energy theory \cite{rhine2019enhanced}, so that even though $H_{\rm sp}$ explicitly breaks time-reversal symmetry, it re-emerges in the low energy theory.

\subsection{Spinon-phonon coupling vertex}\label{sec:spinon_phonon_coupling_cont}
In this section, we will describe a general framework for deriving the spinon-phonon interaction Hamiltonian from symmetry considerations and then apply it to our model, Eq.~(\ref{eq:lowenergyspinon}). Approaches based on symmetry have also been used to find the phonon couplings in graphene \cite{manes2007,basko2009}, but the main difference in our spin-liquid system is that the analysis needs to account for the projective symmetry group of our ansatz. A universal procedure that does exactly this is provided by \citet{serbynlee2013}, and we will reproduce their method here to provide background.

We begin by specifying the form of the spinon-phonon interaction Hamiltonian,
\begin{equation}
    \mathcal{H}^{}_{\rm sp-ph}=\int\dfrac{d^2\vec{k}\,d^2\vec{q}}{(2\pi)^4}\Psi_a^\dagger(\vec{k+q}){h}^{}_{\rm sp-ph}(\vec{k},\vec{q})\Psi^{}_a(\vec{k}).
\end{equation}
Expanding $\vec{k}$ around the Dirac points at $\Gamma$, we allow the presence of terms of zeroth order, ${h}^{(0)}_{\rm sp-ph}(\vec{q})$, and linear order, ${h}^{(1)}_{\rm sp-ph}(\vec{k},\vec{q})$, in the spinon momentum $\vec{k}$ so that the total interaction Hamiltonian can be written as 
\begin{equation}
{h}^{}_{\rm sp-ph}(\vec{k},\vec{q})={h}^{(0)}_{\rm sp-ph}(\vec{q})+{h}^{(1)}_{\rm sp-ph}(\vec{k},\vec{q}).
\end{equation}
Often, only the zeroth-order contribution ${h}^{(0)}_{\rm sp-ph}(\vec{q})$ needs to be considered, but as we will find for the nonchiral $\pi$-flux state, ${h}^{(0)}_{\rm sp-ph}(\vec{q})=0$ by symmetry. In the case of nonzero $t_2$, there is a single symmetry-allowed zeroth-order phonon coupling. Either way, to obtain a nonzero $\eta^H$, it will be necessary to also take ${h}^{(1)}_{\rm sp-ph}(\vec{k},\vec{q})$ into account. The ${h}^{(1)}_{\rm sp-ph}$ term can be understood as a deformation of the spinon band structure near the Dirac points at $\Gamma$ by acoustic phonons.

Acoustic phonons can only couple to the spinons through spatial derivatives of the phonon field, so they enter into ${h}_{\rm sp-ph}(\vec{k},\vec{q})$ through the $\vec{q}$ Fourier component of $\vec{u}(\vec{r})$. As in the previous section, we expect couplings to the time derivative of $\vec{u}$ to be suppressed by the ratio of the sound and Fermi velocities, so we will ignore them in our analysis. Since both the phonon fields and the phonon momenta transform under the vector representation $E_1$ of $C_{4v}$, we can decompose the set of terms $\partial_iu_j(\vec{r})\sim-iq_iu_j(\vec{q})$ into irreducible representations as
\begin{equation}
\label{eq:symm_phon}
E_1^{\rm ph}\otimes E_1^{\rm ph}=\oplus^{}_j D_j^{\rm ph},
\end{equation}
where $D_j^{\rm ph}$ labels irreducible representations of $C_{4v}$. As spinons are fermionic while phonons are bosonic, the leading-order coupling of phonons must be to bilinears of the continuum spinon fields $\psi$. Even though $\psi$ realizes a projective representation of the lattice symmetry group $C_{4v}'$, the space of local spinon bilinears, 
\begin{equation}
  G^{}_{\psi^\dagger\psi}=\left\{  \psi^\dagger\mathbb{I}\psi,\;\psi^\dagger\tau^i\psi,\;\psi^\dagger\mu^i\psi,\;\psi^\dagger(\mu^i\tau^j)\psi\right\},
\end{equation}
realizes regular representations in our Abelian U(1) spin liquid because the U(1) gauge factors cancel. For non-Abelian SU(2) spin liquids, as studied in Refs.~\onlinecite{hermele2004,hermele2005,hermele2008kagome}, we must restrict ourselves to spin singlet bilinears to obtain regular representations. We omit the spin degrees of freedom since we assume that the phonons couple to both $\psi_{\uparrow}$ and $\psi_{\downarrow}$ bilinears equally. In similar fashion to the phonons, we can decompose the representation of all bilinears into irreducible representations,
\begin{equation}
\label{eq:symm_spinon_0}
   G^{}_{\psi^\dagger\psi}=\oplus^{}_j D_j^{\psi^\dagger\psi}. 
\end{equation}
As $h^{(1)}_{\rm sp-ph}$ includes terms that couple spinon momenta $\vec{k}$ (transforming in the vector representation) and bilinears, we must also consider
\begin{equation}
\label{eq:symm_spinon_1}
E_1^{\rm sp}\otimes G^{}_{\psi^\dagger\psi}=\oplus^{}_j D_j^{\vec{k},\psi^\dagger\psi}.
\end{equation}
We observe that $\mathcal{H}_{\rm sp-ph}$, which could possibly contain terms like
\begin{align}
    \sum_{ij} \left(D_i^{\rm ph}\otimes D_j^{\psi^\dagger\psi}+D_i^{\rm ph}\otimes D_j^{\vec{k},\psi^\dagger\psi}\right)
\end{align}
must be invariant under all symmetries. This is only possible if the representations are equal; that is, $D_i^{\rm ph}= D_j^{\psi^\dagger\psi}$ or $D_i^{\rm ph}= D_j^{\vec{k},\psi^\dagger\psi}$. Therefore, pairing together basis functions of equivalent irreducible representations between Eqs.~(\ref{eq:symm_phon}) and  (\ref{eq:symm_spinon_0}) and (\ref{eq:symm_spinon_1}) will give us all possible couplings in $h_{\rm sp-ph}$. Furthermore, the additional SU(2) symmetries of time-reversal and charge conjugation will impose further constraints on allowed couplings, as the phonon strain field $\partial_iu_j$ is invariant under both symmetries.

Applying this formalism to our lattice symmetry group $C_{4v}$, the underlying symmetry group of the phonons, we have 
\begin{equation}
\label{eq:phonon_reps}
    \oplus^{}_j D_j^{\rm ph}=A^{}_1\oplus A^{}_2\oplus B^{}_1\oplus B^{}_2,
\end{equation}
in \equref{eq:symm_phon}, with basis elements $\partial_xu_x+\partial_yu_y,$ $\partial_x u_y-\partial_yu_x,$ $\partial_x u_x-\partial_y u_y,$ and $\partial_xu_y+\partial_yu_x$, respectively.
For the spinon sector, we can decompose the bilinears into representations of $C_{4v}'$,
\begin{alignat}{2}
G^{}_{\psi^\dagger\psi}&=A^{}_1\oplus A^{}_2\oplus \cdots,\\
\label{eq:spinon_reps}
E_1^{\rm sp}\otimes G^{}_{\psi^\dagger\psi}&=A^{}_1\oplus A^{}_2\oplus B^{}_1\oplus B^{}_2\oplus\cdots,
\end{alignat} where the ellipsis stands for irreducible representations of $C_{4v}'$ that transform nontrivially under lattice translations; these cannot be coupled to the phonons, which transform trivially under translations. The full results and explicit basis elements are tabulated in Section III of Ref.~\onlinecite{serbynlee2013}.

The ostensibly allowed couplings between the $(A_1,A_2)$ components in $G_{\psi^\dagger\psi}$ and $E_1^{\rm ph}\otimes E_1^{\rm ph}$ turn out to be forbidden by time-reversal symmetry, as the $(A_1,A_2)$ components in $G_{\psi^\dagger\psi}$,
\begin{equation}
D_{A_1}^{\psi^\dagger\psi}=\left\{\psi^\dagger\mathbb{I}\psi\right\},\quad
  D_{A_2}^{\psi^\dagger\psi}=\left\{\psi^\dagger\tau^z \psi\right\},
\end{equation}
are both time-reversal odd. However, there is an allowed coupling to the $A_2$ channel through the orbital current $t_2$. Since the orbital current $t_2$ also transforms as $A_2$, by coupling $t_2\propto (m+M)$ and $D_{A_2}^{\psi^\dagger\psi}$ together, we obtain a term that transforms trivially (as $A_1$). This term is permitted because $D_{A_2}^{\psi^\dagger\psi}$, like $t_2$, is time-reversal odd, so the product $t_2\cdot D_{A_2}^{\psi^\dagger\psi}$ can couple to the phonon density fluctuations. Therefore, we find 
\begin{equation}
\label{eq:int_ham_cont_h0}
    h_{\rm sp-ph}^{(0)}(\vec{q})=  ig^{}_0 \qty(m+M) \tau^z \qty(q^{}_xu^{}_x+q^{}_yu^{}_y),
\end{equation}
with $g_0$ labeling some phenomenological coupling coefficient.
Note that we cannot couple $m\tau^z$ to phonons as $m$ itself is not an irreducible representation (it includes the N\'eel order), but the combination $m+M=t_2$ is irreducible. 

The bilinears in Eq.~(\ref{eq:spinon_reps}) suffer no such restriction as they are all time-reversal and charge-conjugation invariant. The basis elements for the irreducible representations in Eq.~(\ref{eq:spinon_reps}) are analogous to those in Eq.~(\ref{eq:phonon_reps}), with the replacement $u_i\rightarrow \tau^i$. Now, we can couple each of the first four irreducible representations in Eq.~(\ref{eq:spinon_reps}) to its partner in Eq.~(\ref{eq:phonon_reps}). For example, the $A_1$ spinor-bilinear component is of the form
\begin{align}
   D_{A_2}^{\vec{k}\psi^\dagger\psi}  =\left\{ \psi^\dagger (k^{}_x \tau^x + k^{}_y \tau^y) \psi\right\},
\end{align}
so that the $A_1$-$A_1$ coupling contribution to ${h}_{\rm sp-ph}$ will be of the form $ig_{A_1}(q_xu_x+q_yu_y)(k_x\tau^x+k_y\tau^y)$ for some coupling constant $g_{A_1}$.
After some simplification, the end result is
\begin{alignat}{2}
\nonumber
\label{eq:int_ham_cont}
{h}^{(1)}_{\rm sp-ph}(\vec{k},\vec{q})=&i(&&g^{}_1 q^{}_x k^{}_x \tau^x + g^{}_2q^{}_y k^{}_y \tau^x+ g^{}_3 q^{}_y k^{}_x \tau^y \\&+ &&g^{}_4 q^{}_x k^{}_y \tau^y)u_x+(x\leftrightarrow y),
\end{alignat} for phenomenological couplings $g_i$. The $g_i$ label combinations of irreducible representations, with $g_{1,4}=g_{A_1}\pm g_{B_1}$ and $g_{2,3}=g_{B_2}\pm g_{A_2}$.

\subsection{Evaluation of phonon polarization and Hall viscosity}
As in Eqs.~(\ref{eq:effective_action}) and (\ref{eq:Lattice_EffAction}), we will now integrate out the fermion fields to obtain the Hall viscosity for the phonon fields. From Eqs.~(\ref{eq:int_ham_cont_h0}) and (\ref{eq:int_ham_cont}) we can define our spinon-phonon coupling vertices to be (rewriting in terms of the low-energy Dirac fields $\Psi$)
\begin{widetext}
\begin{subequations}
\begin{alignat}{2}
\mathcal{H}^{}_{\rm sp-ph}&=\int\dfrac{d^2\vec{k}\,d^2\vec{q}}{(2\pi)^4}\Psi_a^\dagger(\vec{k+q})\lambda_{k,q}^iu^{}_i\Psi^{}_a(\vec{k}),\\\label{eq:vertex_contx}
\lambda_{\vec{k},\vec{q}}^x&=i(g^{}_1 q^{}_x k^{}_x \tau^x + g^{}_2q^{}_y k^{}_y \tau^x+ g^{}_3 q^{}_y k^{}_x \tau^y + g^{}_4 q^{}_x k^{}_y \tau^y + g^{}_0 q^{}_x (m+M) \tau^z),\\
\label{eq:vertex_conty}\lambda_{\vec{k},\vec{q}}^y&=i(g^{}_1 q^{}_y k^{}_y \tau^y + g^{}_2q^{}_x k^{}_x \tau^y+ g^{}_3 q^{}_x k^{}_y \tau^x + g^{}_4 q^{}_y k^{}_x \tau^x + g^{}_0 q^{}_y  (m+M) \tau^z).
\end{alignat}
\end{subequations}
The last term coming from the coupling of the orbital current $t_2=m+M$ in $\lambda_{k,q}^{x,y}$ has no dependence on spinon momentum $\vec{k}$. 
We can write the phonon self-energy, as in Fig.~\ref{fig:feynman_diagram}(b), in Matsubara frequency space as
\begin{equation}\label{eq:selfenergy_cont}
\Pi^{xy}(\vec{q},i\Omega_m)=-\dfrac{1}{2}\int_{\vec{k},\omega^{}_n}2\cdot\mathrm{Tr}\left[\lambda^y_{\vec{k},\vec{q}}G(\vec{k},i\omega_n)\lambda^x_{\vec{k}+\vec{q},-\vec{q}}G(\vec{k}+\vec{q},i\omega^{}_n+i\Omega^{}_m)\right],
\end{equation}
\end{widetext}where $G(\vec{k},i\omega_n)$ denotes the Dirac fermion Green's function,
\begin{equation}
      G(k,\omega)
      =\dfrac{\omega\mathbb{I}+\mathbf{H}_{\vec{k}}\cdot\boldsymbol{\tau}}{\omega^2-\mathbf{H}_{\vec{k}}^2},\quad \mathbf{H}_{\vec{k}}\equiv (k_x,k_y,m).
\end{equation} In Eq.~(\ref{eq:selfenergy_cont}), we define $\int_{k,\omega_n}$\,$\equiv$\,$ T\sum_{\omega_n}\int d^2\vec{k}/(2\pi)^2$, and we have also included a factor of $2$  to account for the two species of Dirac fermions. The Hall viscosity originates from the off-diagonal, antisymmetric component of $\Pi^{xy}$. In real frequency, Eq.~(\ref{eq:selfenergy_cont}) contributes a term to the phonon effective action of the form
\begin{equation}
    \delta S^{}_{\rm{eff}}=\int_{\vec{q},\Omega}\Pi^{\mu\nu}(\vec{q},\Omega)u^{}_\mu(-\vec{q},-\Omega) u^{}_{\nu}(\vec{q},\Omega)
\end{equation}
from which we can extract  
\begin{align} \label{eq:continuum_etaH_selfenergy}
\eta^H&=\lim_{q\rightarrow0}\lim_{\Omega\rightarrow 0}-\dfrac{1}{\Omega L^d}\partial^2_{\vec{q}}\Im\left[\Pi^{xy}(\vec{q},\Omega)\right]
\end{align}
In our continuum model, we did not consider any coupling to the rotational strain field, so $\eta^M=0$. Other terms in the phonon self-energy, such as the diagonal and symmetric components, will renormalize the real part of the phonon propagator. This is a small effect that does not contribute to phonon chirality, so we will not consider it here. As we are only interested in the leading-order contributions of $q$ and $\Omega$, we use \begin{equation}
\label{eq:vertex_approx_cont}
    \lambda^\mu_{\vec{k}+\vec{q},-\vec{q}}=\lambda^\mu_{\vec{k},-\vec{q}}+\mathcal{O}(\vec{q}^2)
\end{equation} and neglect the anharmonic contributions. Relegating the details of the computation to Appendix~\ref{app:selfenergy_continuum}, we find that
\begin{widetext}
\begin{align}\nonumber
\Pi^{xy}(\vec{q},i\Omega_m)&=\vec{q}^2(g^{}_1g^{}_2-g^{}_3g^{}_4)\dfrac{m\Omega}{8\pi}\left(\Lambda-2|m|\right)+\vec{q}^2g^{}_0(g^{}_2-g^{}_3)(m+M)\dfrac{\Omega}{8\pi}\left(\Lambda-2|m|\right),
\end{align} 
where $\Lambda$ is a UV cutoff near the Dirac points. Then, Eq.~\eqref{eq:continuum_etaH_selfenergy} gives us 
\begin{align}\label{eq:etah_continuum}
\eta^H&=\dfrac{1}{4\pi L^2}\left[(g^{}_1g^{}_2-g^{}_3g^{}_4)m+g^{}_0(g^{}_2-g^{}_3)(m+M)\right](\Lambda-2|m|),
\end{align}
\end{widetext}
 after analytic continuation to real $\Omega$. To compare the continuum result to the lattice, we need to extract the leading nonanalytic contribution:
\begin{align}
    \eta^H \sim -\qty[g^{}_0(g^{}_2-g^{}_3)(m+M)]|m|\sim M
    |m|;
\end{align}
we see that the second derivative $\partial^2{\eta^H}/\partial^2{m}$\,$\propto$\,$ \delta(m)$ is divergent in the limit $m\rightarrow 0$, in agreement with what we found in Eq.~\eqref{eq:Lattice_HallVisc_CrPtDeriv} on the lattice.

From Eq.~\eqref{eq:etah_continuum}, we notice that the Hall viscosity $\eta^H$ scales with the two effective couplings $g_1g_2-g_3g_4$ and $g_0(g_2-g_3)$. This can be understood in the representation theory framework presented earlier, as both $g_1g_2-g_3g_4$ and $g_0(g_2-g_3)$ transform in the $A_2$ channel of $C_{4v}$, which descends to the $A_1$ channel of $C_4$ as reflection symmetry is broken in our ansatz. Further discussions on this point are included at the end of Appendix \ref{app:selfenergy_continuum}.

\subsection{Discussion and comparison to the lattice results}\label{sec:continuum_disc}
Our analysis highlights that the lattice symmetries strongly constrain the allowed spinon-phonon couplings. Therefore, though most spin-liquid phases of interest have similar Dirac excitations and effective theories in the continuum, the allowed spinon-phonon interactions and resulting Hall viscosity $\eta^H$ in the continuum are sensitive to microscopic information about the phase.

We contrast our result with the quantized Hall viscosity found in Refs.~\onlinecite{rosch2018,yeperkins2020} for  Majorana fermions in the gapless B phase of the Kitaev honeycomb model \cite{kitaev2006anyons}. This is a special feature of the lattice symmetry group $C_{6v}$, as in addition to a trivial density fluctuation coupling, the zero-flux phase \cite{basko2009,serbynlee2013} on a honeycomb lattice allows a spinon-phonon interaction in the $E_2$ channel of the form 
\begin{equation}
    h_{\rm{sp-ph}}^{(0)}(\vec{q})\sim \left[(q^{}_xu^{}_x-q^{}_yu^{}_y)\tau^x-(q^{}_xu^{}_y+q^{}_yu^{}_x)\tau^y\right]\mu^z
\end{equation}
to zeroth order in the spinon momentum $\vec{k}$ near the Dirac point, where $\vec{q}$ is the phonon momentum. This additional coupling, in which the spinon momentum $\vec{k}$ does not appear, should be understood as a consequence of the special symmetries of the honeycomb lattice. Integrating out the spinons on the honeycomb lattice then leads to a discontinuous Hall viscosity \cite{rosch2018,yeperkins2020},
\begin{equation}
   \eta^H\sim \mathrm{sign}(m),
\end{equation}
that depends only on the sign of the Dirac mass $m$. Moreover, it was found in Ref.~\onlinecite{yeperkins2020} that for the Kitaev spin liquid, $\eta^H$ decreased as the magnitude of the time-reversal-symmetry-breaking perturbation increased.

In our analysis for the square lattice, we see that the nonanalytic behavior of the continuum $\eta^H$ agrees with the lattice result, Eq.~\eqref{eq:Lattice_EtaH}, at low energies and near the Dirac point. However, it should be noted that the continuum viscosity is in general regularization-dependent, and only the difference in $\eta^H$ between two phases is universal \cite{hughesfradkin2011,hughes2013long}. With this in mind, we observe that the difference in $\eta^H$ across the QPT scales, at leading order, linearly with the Dirac mass $m=2t_2-N/4$ in both the lattice and continuum formulations. This differs from the Hall viscosity obtained for the Dirac Chern insulator on a square lattice \cite{shapourian2015viscoelastic} in which case, the Hall viscosity of the tight-binding Hamiltonian scales quadratically with the Dirac mass. For the continuum Dirac field theory of the Chern insulator, introducing suitable Pauli-Villars regulators and counterterms eliminates the dependence of the viscosity on the UV cutoff $\Lambda$ and also leads to a quadratic dependence of $\eta^H$ on $m$ \cite{hughes2013long}.

In our chiral spin liquid ansatz in Eq.~(\ref{eq:spinon_lattice}), the phonons are coupled directly to the orbital current $t_2$ but not the effective Dirac mass $m=2t_2-N/4$. This is caused by the presence of N\'eel order, which does not couple to lattice distortions as it is purely an on-site term [see \equref{eq:spinon_lattice}]. In the limit $N\rightarrow 0$, the two Dirac masses coincide and for our calculation on the lattice, $\eta^H\sim M|m|= m|m|$, as in the case of the Chern insulator. In the continuum, we reproduce the field theory of the Chern insulator as in Refs.~\onlinecite{hughesfradkin2011,hughes2013long}, and with further regularization, the same scaling is obtained for the Hall viscosity. In both the lattice and the continuum, the divergence in $\eta^H$ is then only visible in its third derivative with respect to $m$.

\begin{figure}[tb]
\includegraphics[width=\linewidth]{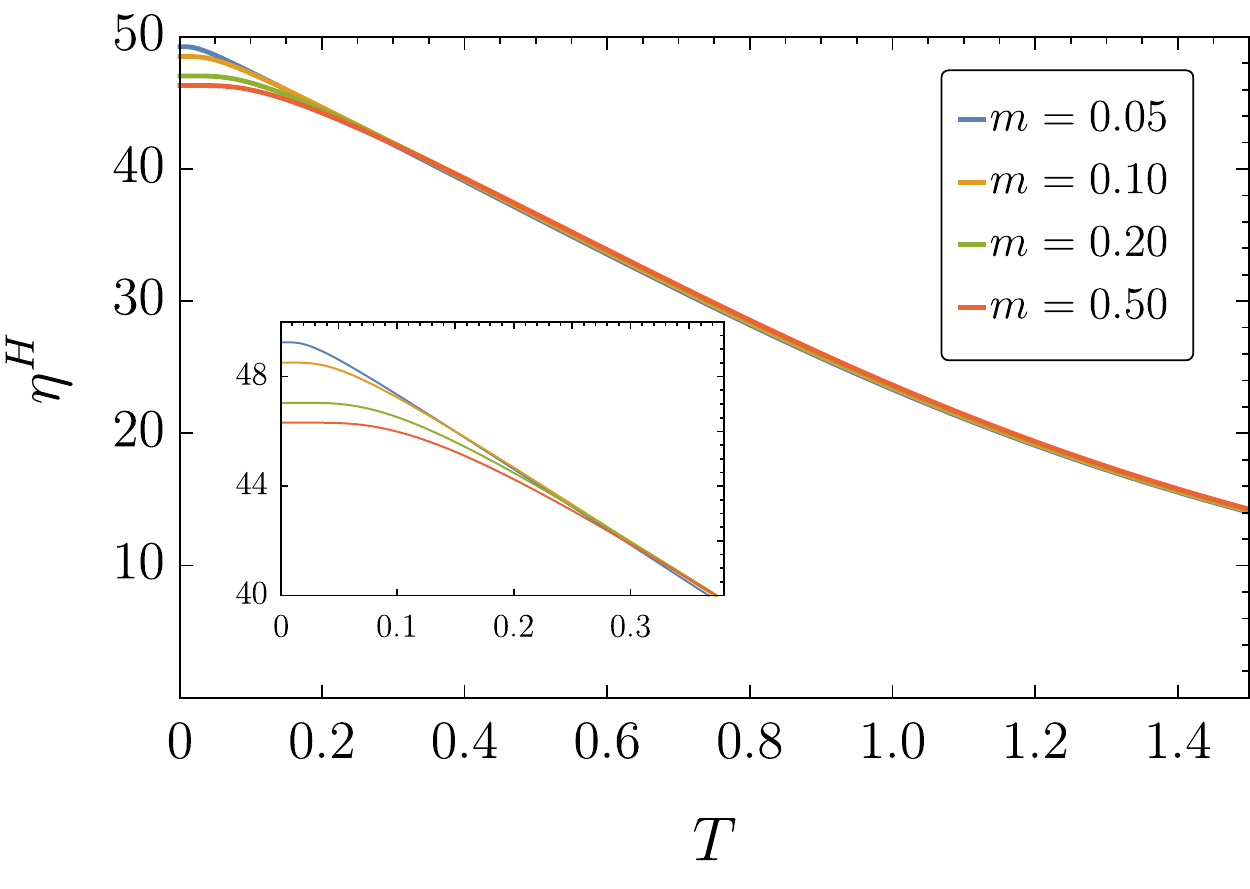}
\caption{Finite-temperature scaling of the continuum phonon Hall viscosity from Eq.~\eqref{eq:etah_finiteTExact}, with $F_1=2F_2=2$ and $M=\Lambda=5$. The inset shows the low-temperature plateaus of $\eta^H$ with a scale set by $m$.}
\label{fig:eta_finiteT}
\end{figure}

Finally, our analysis can be extended to include the finite-temperature result [see \equref{eq:etah_continuum_appendix} in Appendix~\ref{app:selfenergy_continuum}], 
\begin{alignat}{1}
\eta^H= (F_1M+F_2m)\cdot &\big[D_\Lambda(T,m)\nonumber\\
&-4T\log(2\cosh(|m|/2T))\big]\label{eq:etah_finiteTExact},
\end{alignat} for some function $D$ dependent on $m$, $T$, and a UV cutoff $\Lambda$. The constants $F_1$ and $F_2$ are combinations of the spinon-phonon couplings. In the limit that $\Lambda\gg m,T$, we have 
\begin{alignat}{1}
\label{eq:etah_finiteT}
\eta^H= (F_1M+F_2m)\cdot &\big[4T\log\left(2\cosh(\Lambda/2T)\right) \\
\nonumber&-\Lambda-4T\log(2\cosh(|m|/2T))\big].
\end{alignat}
As the hyperbolic cosine is an even and positive function, we find that the viscosity is smooth at finite $T$; this is expected because the Matsubara summation, at finite temperatures, does not introduce any nonanalyticities. In the limit of $M$\,$\gg$\,$ m$, we see that the $\Lambda$-independent part of $\eta^H/(M T)$ is only a function of the ratio $|m|/T$. 
The temperature and $m$ dependence of $\eta^H$ arising from \equref{eq:etah_finiteTExact} is illustrated in Fig.~\ref{fig:eta_finiteT}. The zero-temperature value of $\eta^H$ depends on the momentum cutoff $\Lambda$. We observe, in particular, that $\eta^H$ decays at high temperature and plateaus near zero temperature, with the size of the plateau dependent on the mass gap $m$. These universal features were also present in our lattice calculation, in Figs.~\ref{fig:Lattice_fig1}(b) and \ref{fig:Lattice_fig1}(d).

\section{Physical Consequences}\label{sec:exp}
For acoustic phonons, the dispersion is assumed to be $\omega_{\rm{ph}}\propto|\vec{q}|+\mathcal{O}(\vec{q}^2)$, so, according to Eq.~\eqref{eq:delta_s_eta}, the Hall viscosity's contribution to the phonon effective action is of order $\partial u\partial\dot{u}\sim |\vec{q}|^3u^2$. This is more relevant than the leading anharmonic correction, which is of order $\vec{q}^4$. Note that the other possible $\mathcal{O}(\vec{q}^3)$ contribution to the phonon action
\begin{equation}
    \int d^2x\,d t \;D^{}_{ijklm}\partial^{}_i\partial^{}_j u^{}_k\partial^{}_l u^{}_m
\end{equation} vanishes in the presence of inversion symmetry. In two-dimensional isotropic systems, it was found that the Hall viscosity mixes the longitudinal and transverse modes and renormalizes the phonon spectrum \cite{barkeshli2012,yeperkins2020}, 
\begin{equation}
   \Delta \omega^{}_{\rm{ph}}\sim \eta^2q^3.
\end{equation}
However, the exact numerical prefactor of the correction, estimated to be very small by \citet{barkeshli2012}, requires knowledge of the energies associated with the appropriate spin-lattice couplings, and the phonon spectrum cannot distinguish the sign of the Hall viscosity. Another consequence of a Hall viscosity is phonon Faraday rotation, which describes the rotation of the linear polarization vector of transverse acoustic phonons due to splitting in the circularly-polarized velocities \cite{wang1971acoustic, sytcheva2010magneto,shapourian2015viscoelastic}.

Recently, the thermal Hall effect has emerged as a powerful probe of neutral excitations such as spinons, prompting extensive experimental and theoretical studies in a variety of correlated quantum materials, including the cuprate superconductors \cite{grissonnanche2019giant, grissonnanche2020chiral, prb2019,han2019consideration,rhine2019enhanced,li2019thermal,chen2020enhanced,haoyu2020gauge}
and Kitaev materials like $\alpha$-RuCl$_3$ \cite{kasahara2018majorana,kasahara2018unusual,hentrich2018large, balz2019finite, yokoi2020half, yamashita2020, ye2018quantization,rosch2018,cookmeyer2018spin,egmoon2020,teng2020unquantized}.
Here, we observe that a phonon Hall viscosity, in general, implies a nonzero phonon thermal Hall conductivity by imparting a Berry curvature to the phonon energy bands. Moreover, their relative signs can be determined given the coupling constants. As previously studied, a phonon thermal Hall response can arise from a coupling of phonons to the magnetization of the system \cite{zhangren2010}. More recently, in ferroelectric insulators \cite{chen2020enhanced}, the flexoelectric coupling of acoustic phonons to the dipole density was shown to lead to a thermal Hall response. In our case, the Hall viscosity appears in the phonon effective action as a term analogous to those flexoelectric couplings. For example, consider isotropic phonons in two dimensions,
\begin{align}
    S^{}_{\rm ph} = \dfrac{1}{2}\int d^2x\,d t\; \rho \dot{\vec{u}}^2+\mu^{}_1 \grad \vec{u}^2 +\mu^{}_2(\grad \cdot\vec{u})^2, \label{eq:discussion_HV_example}
 \end{align} with mass density $\rho$ and elastic constants $\mu_i$.
Assuming the Hall viscosity in Eq.~\eqref{eq:Intro_hallViscosityLagran}, the thermal Hall conductivity \cite{qin2012}, in the low-temperature limit, reads
\begin{alignat}{2}
    \kappa_{xy}^{2d}(T)&=-\eta \dfrac{3 \zeta(3)k_B^3}{\pi \hbar^2}  \left[\frac{g_1}{\sqrt{\mu_1}} - \frac{g_2}{\sqrt{\mu_1+\mu_2}}\right]T^2,\\
    \nonumber g_1&\equiv \frac{4 \mu_1 + \mu_2}{2 \sqrt{\mu_1}\mu_2}; \quad g_2\equiv\frac{4 \mu_1 + 3\mu_2}{2 \mu_2 \sqrt{\mu_1 + \mu_2}}.
\end{alignat}
Hence, given that $\eta$ plateaus at low temperature from our analyses in Secs.~\ref{sec:lattice_phonon_bubble} and \ref{sec:continuum_disc}, we see that $\kappa_{xy}^{2d}/T \propto T$ as $T\rightarrow0$ (note that we expect that $\kappa_{xy}^{3d}/T \propto T^2$ at low temperature \cite{qin2012}). While we are unable to make quantitative estimates of the strength of this response, we have demonstrated that this effect generically exists for both the conventional N\'eel phase and the N\'eel state coexisting with semion topological order in \figref{fig:PD}, providing an intrinsic source of phonon chirality. With an eye towards recent experiments on the phonon thermal Hall response in cuprates \cite{grissonnanche2020chiral,boulanger2020thermal}, our proposal lays the foundation for work on possible enhancement of heat transport due to extrinsic mechanisms in these topological systems \cite{haoyu_subir}. 

\section{Conclusion and Outlook}\label{sec:conclusion}
In this paper, we have analyzed the phonon Hall viscosity arising from the coupling to spin degrees of freedom on the square lattice in a magnetic field.

We employed a fermionic spinon formulation and obtained a low-energy effective action for the phonon fields by integrating out the spinons. Two complementary approaches were studied: first, starting from the lattice spinon model of Ref.~\onlinecite{rhine2019enhanced}, we introduced the coupling to lattice vibrations using the physical model of bond stretching (or equivalently, adiabatic response). In the second approach, only the relevant low-energy spinon degrees of freedom were retained, and the resulting continuum Dirac theory was coupled to lattice vibrations purely by symmetry considerations. 

Even in the continuum limit, microscopic details about the lattice symmetry were shown to have drastic effects on the critical behavior of $\eta^H$: as opposed to the discontinuity of $\eta^H$ when changing the sign of the effective Dirac mass $m$ at the transition on the honeycomb lattice, we demonstrated that the symmetries of the square lattice lead to a Hall viscosity that varies linearly with the effective Dirac mass $m$, in both the continuum and the lattice theory. We also calculated $\eta^H$ at finite temperature and determined a scaling form for the ratio $\eta^H/(TM)$.

The Hall viscosity is a measure of time-reversal symmetry breaking in the spinon sector, and its nonanalyticities can serve as signatures of the field-driven topological quantum phase transition. 
We found that the second derivative of the Hall viscosity with respect to the mass, $\partial_m^2\eta^H$, diverges at the transition ($m=0$) between the two phases in \figref{fig:PD} at zero temperature. This leads to a kink in the field dependence of $\eta^H$ [see Figs. \ref{fig:Lattice_fig1}(a) and \ref{fig:Lattice_fig1}(c)]. We showed that this enhanced singular behavior---as compared to the square-lattice Chern insulator where $\partial_m^3\eta^H$ diverges \cite{shapourian2015viscoelastic}, can be traced back to the presence of N\'eel order.

In addition to previous studies which have proposed measuring the Hall viscosity through various phononic properties of the material, we have shown how the Hall viscosity also leads to an intrinsic thermal Hall response. This response can potentially be enhanced by extrinsic scattering mechanisms and may be detectable in experiments which indicate that phonons are the dominant contribution to heat transport. As more and more experiments probe the exotic nature of topological phases and possible spin liquid candidates, we believe that the phonon Hall viscosity can be a powerful tool for detecting fractionalization and quantum critical phenomena.

Finally, we note that our computations were carried out in the setting of a spinon mean-field theory. Gauge fluctuations can potentially change the spinon phase diagram as well as renormalize the spinon-phonon couplings \cite{haoyu2020gauge}. However, we expect the nonanalytic signature of a quantum phase transition to persist. The exact effects of gauge-field fluctuations on phonon dynamics remain an open question, and their consequences are avenues for further study.

\subsection*{Acknowledgements}
We thank Haoyu Guo for collaborations on related projects \cite{rhine2019enhanced,haoyu_subir,haoyu2020gauge}, and him and Ga{\"e}l Grissonnanche, Natalia Perkins, and Louis Taillefer for valuable discussions. This research was supported by the National Science Foundation under Grant No.~DMR-2002850. This work was also supported by the Simons Collaboration on Ultra-Quantum Matter, which is a grant from the Simons Foundation (651440, S.S.).\\
Y.Z. and Y.T. contributed equally to this work.\\
{\it Note added:} Recently, we learned of the related work in Ref.~\onlinecite{Mengxing21}.

\appendix
\section{Hall viscosity in the absence of a Zeeman field}\label{app:EtaH_derivation}
To evaluate the effective action, we have to compute  (using $\Tr[\tau^\alpha\tau^\beta\tau^\delta]=2i\varepsilon^{\alpha\beta\delta}$, $\varepsilon$ being the Levi-Civita tensor) the trace
\begin{widetext}
\begin{equation}
\Tr \qty[\tau^\beta G (\vec{k},i\omega^{}_n)\tau^\alpha G (\vec{k},i(\omega^{}_n+\Omega^{}_m))]=\sum_{\pm}\dfrac{2\varepsilon^{\alpha\beta\delta}\Omega^{}_m \mathbf{H}_{\pm,\vec{k}}^\delta}{(\mathbf{H}_{\pm,\vec{k}}^2+\omega_n^2)(\mathbf{H}_{\pm,\vec{k}}^2+(\omega^{}_n+\Omega^{}_m)^2)}=\sum_{\pm}\dfrac{2\varepsilon^{\alpha\beta\delta}\Omega^{}_m \mathbf{H}_{\pm,\vec{k}}^\delta}{(\mathbf{H}_{\pm,\vec{k}}^2+\omega_n^2)^2}+\mathcal{O}(\Omega_m^2),
\end{equation}
where we have only kept the terms antisymmetric and linear in $\Omega$ since these are the only ones that will contribute to the Hall viscosity.
This approximation is valid because we are only extracting the first-order (in $\Omega$) contribution to the effective action. The Matsubara summation yields
\begin{align}
T\sum_{\omega_n}\dfrac{1}{(\mathbf{H}_{\pm,\vec{k}}^2+\omega_n^2)^2}=\dfrac{1-2n^{}_F(\abs{\mathbf{H}^{}_{\pm,\vec{k}}})+2\abs{\mathbf{H}^{}_{\pm,\vec{k}}}n_F'(\abs{\mathbf{H}^{}_{\pm,\vec{k}}})}{4\abs{\mathbf{H}^{}_{\pm,\vec{k}}}^3}.
\end{align} Therefore, we can extract from 
\begin{align}S^{}_{\text{eff}}&= - \frac{1}{2 L^2 \beta^2} \sum_{ \omega_n, \Omega_m, \vec{k}} \Tr \qty[K^{}_{\vec{k},i\Omega_m}G (\vec{k},i\omega^{}_n)K^{}_{\vec{k},-i\Omega_m}G (\vec{k},i(\omega^{}_n+\Omega^{}_m))]\end{align} 
the Hall viscosity 
\begin{equation}
    \eta^H=-\dfrac{1}{L^2}\sum_{\alpha\beta\delta}\sum_{\vec{k},\pm} \varepsilon^{\alpha\beta\delta}\gamma^{\alpha}_{xx}\gamma_{xy}^{\beta}\mathbf{H}_{\pm,\vec{k}}^\delta\cdot\dfrac{1-2n^{}_F(\abs{\mathbf{H}^{}_{\pm,\vec{k}}})+2\abs{\mathbf{H}^{}_{\pm,\vec{k}}}n_F'(\abs{\mathbf{H}^{}_{\pm,\vec{k}}})}{4\abs{\mathbf{H}^{}_{\pm,\vec{k}}}^3}.
\end{equation}
Writing out the summation over $\alpha,\beta,\delta$ explicitly leads to Eq.~(\ref{eq:Lattice_EtaH}) of the main text.
\end{widetext}

\section{Rotational strain field coupling}\label{app:EtaM_derivation}
The rotational strain field $\theta_{ij} = \qty(\partial_i u_j - \partial_j u_i)/2$, representing a vorticity, is ordinarily not considered in geometric treatments of phonon interactions. However, given the symmetries of our ansatz, we can couple the time derivative of $\partial{\theta_{ij}}/\partial{t}$ to the orbital current $t_2$ \cite{scheurer_sachdev_2018, sen1995large}:
\begin{align}
        |t^{}_{\vec{n},\vec{n}\pm(\vec{x}+\vec{y})}| = |t^{}_{\vec{n},\vec{n}\pm(\vec{x}-\vec{y})}| \approx t^{}_2 +\lambda^{}_3 \pdv{\theta^{}_{xy}}{t}. 
\end{align}
Since the rotational strain field has only one polarization $\theta_{xy}$, this gives an additional interaction term in \equref{eq: Lattice_bondStretching_couplingVertex} that couples to $\dot{\theta}_{xy}$,
\begin{align}
       \tilde{\gamma} _{\mu\nu}^z(\Omega) \tau^z = 4 \Omega \lambda^{}_3 \sin(k^{}_x)\cos(k^{}_y)  \tau^z, \quad \mu\nu = xy.
\end{align}
This additional coupling is interesting to consider since it can, in principle, lead to a finite Hall viscosity. Following the same procedure for calculating $\eta^H$, we state the analytic answer for $\eta^M$ below:
    \begin{widetext}
    \begin{align}
         \eta^M  &= \frac{1}{L^2} \sum_{\vec{k}, \pm} \tilde{\gamma}_{xy}^z \Bigg[ \frac{ 2 \mathbf{H}^x_{\pm,\vec{k}} \mathbf{H}^z_{\pm,\vec{k}} \gamma_{xx}^x + \qty(-(\mathbf{H}^x_{\pm,\vec{k}})^2 - (\mathbf{H}^y_{\pm,\vec{k}})^2 + (\mathbf{H}^z_{\pm,\vec{k}})^2) \gamma_{xx}^z}{4 \abs{\mathbf{H}^{}_{\pm,\vec{k}}}^3} \qty[1-2n^{}_F(\abs{\mathbf{H}^{}_{\pm,\vec{k}}})+2\mathbf{H}^{}_{\pm,\vec{k}}n_F'(\abs{\mathbf{H}^{}_{\pm,\vec{k}}})] \nonumber \\
         &+   \frac{\gamma_{xx}^z}{4 \abs{\mathbf{H}^{}_{\pm, \vec{k}}}} \qty[-1+2n^{}_F(\abs{\mathbf{H}^{}_{\pm,\vec{k}}})+2\abs{\mathbf{H}^{}_{\pm,\vec{k}}}n_F'(\abs{\mathbf{H}^{}_{\pm,\vec{k}}})]  \Bigg].\label{eq:Lattice_EtaM}
    \end{align}
     \end{widetext}
This expression simplifies at $T=0$ to $\eta^M =$
\begin{align}
     \frac{1}{L^2} \sum_{\vec{k}, \pm} 2 \tilde{\gamma}_{xy}^z \frac{ \mathbf{H}^x_{\pm,\vec{k}} \mathbf{H}^z_{\pm,\vec{k}} \gamma_{xx}^x - \qty((\mathbf{H}^x_{\pm,\vec{k}})^2 + (\mathbf{H}^y_{\pm,\vec{k}})^2 ) \gamma_{xx}^z}{ \abs{\mathbf{H}^{}_{\pm, \vec{k}}}^3}. \label{eq:Lattice_EtaM0T}
\end{align}
At the critical point, where $m_1=0$, expanding for momenta $\vec{q}$ near the Dirac point at $\mathbf{Q}=\qty(\frac{\pi}{2}, 0)$, we have
    \begin{align}
        \eval{\eta^M}_{m_1 = 0}  \sim\sum_{\vec{q}} \frac{m_2^2}{\abs{\vec{q}}}.
\end{align}
Although the above term seems to have a singularity, it is remedied by the integration measure $d^2{\vec{q}}\sim \abs{\vec{q}}d{\abs{\vec{q}}}$. Therefore, $\eta^M$ goes to a finite value as $\vec{q}\rightarrow 0$. Just as for $\eta^H$, we can analyze $\eta^M$ near the critical point. As we tune towards the QPT, the second derivative of $\eta^M$ is $\delta$-function divergent,
\begin{align}
  \eval{\pdv[2]{\eta^M}{m_1}}_{m_1 = 0} \sim \sum_{\vec{q}} \pdv[2]{}{m_1}\frac{q^2}{ \abs{\mathbf{H}^{}_{+,\vec{k}}}^3}\sim  \partial_{m_1}^2|m^{}_1| \xrightarrow[m^{}_1 \rightarrow 0]{} \infty,
\end{align}
with essentially the same behavior as $\eta^H$.

\section{Hamiltonian and projective symmetry of the chiral $\pi$-flux state}
\label{app:psg_continuum}

From the couplings given by Eqs.~(\ref{eq:continuum_t1}, \ref{eq:continuum_t2_1}, \ref{eq:continuum_t2_2}), we obtain the Hamiltonian for the gauge-transformed ansatz in momentum space as (labeling the sublattices by indices $m,n$)
\begin{align}
    H^{}_{\rm t.b.}=-\sum_{\vec{k},\sigma}f_{\vec{k}m\sigma}^\dagger h^{}_{mn}(\vec{k},\sigma)f^{}_{\vec{k}n\sigma},
\end{align}
where the $4\times 4$ matrix $h$ is given by 
\begin{align}
    h(\vec{k},\sigma)=it^{}_1\left(
\begin{array}{cccc}
	0 & -1+K_1^* & 0 & -1+K_2^*\\
	1-K^{}_1 & 0 & -1-K_2^* & 0\\ 
	0 & -1+K^{}_2 & 0 & 1-K^{}_1\\
	1-K^{}_2 & 0 & -1+K_1^* & 0
\end{array}
\right) \nonumber \\+it^{}_2\left(
\begin{array}{cccc}
	0 & 0 &  \tilde{K}^{}_{A} & 0\\
	0 & 0 & 0 & \tilde{K}^{}_{B} \\ 
	- \tilde{K}_{A}^* & 0 & 0 & 0\\
	0 & -\tilde{K}_{B}^* & 0 & 0
\end{array}
\right) +\dfrac{N\sigma }{2}\left(
\begin{array}{cccc}
	1 & 0 & 0 & 0\\
	0 & -1 & 0 & 0\\ 
	0 & 0 & 1 & 0\\
	0 & 0 & 0 & -1
\end{array}
\right). \label{eq:continuumH_full}
\end{align}
For the equations above, we have defined
\begin{align}
    K^{}_{1,2}&\equiv e^{i\vec{k}\cdot \vec{a}_{1,2}},\\
    \tilde{K}^{}_{A}&\equiv -1-K_1^*-K_2^*-K_1^*K_2^*,\\
    \tilde{K}^{}_{B}&\equiv 1+K^{}_1+K_2^*+K^{}_1K_2^*.
\end{align}
Expanding the Hamiltonian $H_{\rm t.b.}$ around the Dirac points at $\Gamma$ then leads to the effective Dirac Hamiltonian given in  Eq.~\eqref{eq:continuum_diracHam}.

To specify the projective symmetry of the continuous spinor fields $\psi_{\alpha\sigma}$ defined in Eq.~\eqref{eq:continuum_fields}, we need to know how the spinons transform under the relevant symmetry group generators. Following closely the analyses of Ref.~\onlinecite{serbynlee2013}, we begin by specifying the projective action of the symmetry operations on the lattice fermions. The symmetries of the $\pi$-flux ansatz, as in Sec.~\ref{sec:continuum}, are generated by translation by $a \hat{\vec{x}}$, $T_x: \vec{r} \rightarrow T_x\vec{r} = \qty(r_x + a, r_y)$; reflection about the $\hat{\vec{x}}$-axis, $R_x$\,$:$\,$ \vec{r}$\,$\rightarrow$\,$R_x\vec{r}$\,$=$\,$\qty(-r_x, r_y)$; and rotation by $\pi/2$, $C_4$\,$:$\,$\vec{r}$\,$\rightarrow$\,$C_4\vec{r}$\,$=$\,$\qty(r_y, -r_x)$; together, these make up the symmetry group $C_{4v}^\prime$. Furthermore, there are two additional SU(2) symmetries of our ansatz, given (in momentum space) by time-reversal $\mathcal{T}: f_{\vec{k}i\sigma} \rightarrow f_{\vec{k}i\sigma }^\dagger$ and charge-conjugation symmetry, $\mathcal{C}: f_{\vec{k}i \sigma} \rightarrow f_{-\vec{k}i -\sigma}^\dagger$. Note that $\mathcal{T}$ is also accompanied by complex conjugation. While $\mathcal{T}$ flips the spin operator $\vec{S}_i=\frac{1}{2}f_i^\dagger\vec{\sigma}f_i$, $\mathcal{C}$ leaves it invariant. To leave the Hamiltonian invariant under these symmetry operations, we may need to supplement the symmetries with additional gauge transformations; hence, the symmetry is implemented \textit{projectively}. For a U(1)-symmetric ansatz, the gauge factors can be conveniently chosen to be $\pm 1$. For example, $T_x$ is implemented as \begin{align}
    T_x\;:\;&f^{}_{\vec{r},1}\rightarrow -f^{}_{T_x^{-1}\vec{r},2},\quad f^{}_{\vec{r},4}\rightarrow f^{}_{T_x^{-1}\vec{r},3}\nonumber    \\&f^{}_{\vec{r},2}\rightarrow -f^{}_{T_x\vec{r},1},\quad f^{}_{\vec{r},3}\rightarrow f^{}_{T_x\vec{r},4}
\end{align}
and the other transformations can be found similarly:
\begin{subequations}
\begin{alignat}{2}
    R_x\;:\;&f_{\vec{r},1,3}\rightarrow f_{R_x\vec{r},2,4},\quad f^{}_{\vec{r},2,4}\rightarrow -f^{}_{R_x\vec{r},1,3}\\
    C_4\;:\;&f^{}_{\vec{r},1}\rightarrow -f^{}_{C_4\vec{r},2},\quad f^{}_{\vec{r},2,3,4}\rightarrow f^{}_{C_4\vec{r},3,4,1}\\
    \mathcal{T}\;:\; &f^{}_{\vec{k} 1,3} \rightarrow f_{\vec{k}1,3}^\dagger,\quad f^{}_{\vec{k} 2,4} \rightarrow f_{\vec{k}2,4}^\dagger\\
    \mathcal{C}\;:\; &f^{}_{\vec{k}n \uparrow} \rightarrow f_{-\vec{k}n\downarrow}^\dagger,\quad f^{}_{\vec{k}n\downarrow} \rightarrow -f_{-\vec{k}n\uparrow}^\dagger
    \end{alignat}
\end{subequations}
From the form of Eq.~\eqref{eq:continuum_fields}, we can now deduce the action of a symmetry generators on the continuous fields:
\begin{subequations}
\begin{alignat}{2}
    T^{}_x &= \mu^y, \\
    R^{}_x &= i \mu^z \tau^y, \\
    C^{}_4 &= \frac{1}{2} \qty(\mu^x + \mu^y)(1+ i \tau^z). 
\end{alignat} 
\end{subequations} For example, under $T_x$, we have $\psi\rightarrow T_x\psi$.
Likewise, for time-reversal and charge-conjugation symmetries, we find
\begin{subequations}
\begin{alignat}{2}
    \mathcal{T} &: \psi \rightarrow -\mu^z \tau^z (\psi^\dagger)^T, \\
    \mathcal{C} &: \psi \rightarrow \sigma^y \mu^x \tau^x (\psi^\dagger)^T. 
\end{alignat}
\end{subequations} With the symmetries now defined, we can determine how the fermion bilinnears split into irreducible representations. Further details, including background on representation theory and the structure of $C_{4v}'$, can be found in Ref.~\onlinecite{serbynlee2013}.

\begin{widetext}
\section{Continuum phonon self-energy}
\label{app:selfenergy_continuum}
In order to find \begin{align}
\Pi^{xy}(\vec{q},i\Omega^{}_m)&=-\dfrac{1}{2}\int_{\vec{k},\omega^{}_n}2\cdot\mathrm{Tr}\left[\lambda^y_{\vec{k},\vec{q}}G(\vec{k},i\omega^{}_n)\lambda^x_{\vec{k}+\vec{q},-\vec{q}}G(\vec{k}+\vec{q},i\omega^{}_n+i\Omega^{}_m)\right]\nonumber
\\&=-\dfrac{1}{2}\int_{\vec{k},\omega_n}2\cdot\mathrm{Tr}\left[\lambda^y_{\vec{k},\vec{q}}G(\vec{k},i\omega^{}_n)\lambda^x_{\vec{k},-\vec{q}}G(\vec{k}+\vec{q},i\omega^{}_n+i\Omega^{}_m)\right]+\mathcal{O}(q^3),
\end{align} 
it will be convenient to first define 
\begin{equation}
\widetilde{\Pi}^{\alpha\beta}_{\gamma\delta\cdots}(\vec{q},i\Omega_m)=\int_{\vec{k},\omega_n}\mathrm{Tr}\left[(k^{}_\gamma k^{}_\delta\cdots)\tau^\alpha G(\vec{k},i\omega^{}_n)\tau^\beta G(\vec{k}+\vec{q},i\omega^{}_n+i\Omega^{}_m)\right]
\end{equation}
because $\Pi^{xy}(\vec{q},i\Omega_m)$ is a linear combination of terms of the form $q^2\widetilde{\Pi}$. As with the vertex in Eq.~(\ref{eq:vertex_approx_cont}), we can make simplifications based on the fact that we are working in the linear-response regime. We only consider terms of order $\Omega_m$ and $\vec{q}^2$ in $\Pi^{xy}$ for the Hall viscosity, so we just have to keep terms of order $\mathcal{O}(\Omega_m^1)$, $\mathcal{O}(q^0)$ in $\widetilde{\Pi}$. 
We can also observe that of the 25 possible contractions of terms between $\lambda^y$ and $\lambda^x$, most will not contribute to the Hall viscosity, either because they will be symmetric or because they contain spinon momentum terms like $k_xk_y$, which vanish after integrating over $\vec{k}$. The end result is that
\begin{equation}
    \Pi^{xy}(\vec{q},i\Omega^{}_m)=\vec{q}^2(g^{}_1g^{}_2-g^{}_3g^{}_4)\widetilde{\Pi}^{21}_{xx}-\vec{q}^2g^{}_0(g^{}_2-g^{}_3)(m+M)\widetilde{\Pi}^{31}_{y}.
\end{equation}
Here, we used $\widetilde{\Pi}^{\alpha\beta}_{xx}=\widetilde{\Pi}^{\alpha\beta}_{yy}$ and $\Tr[\tau^\alpha\tau^\beta\tau^\gamma]=2i\varepsilon^{\alpha\beta\gamma}$ to simplify the result. Including only the antisymmetric terms (under $2\leftrightarrow 1$) and terms of order $\mathcal{O}(\Omega_m^1q^0)$, 
we can evaluate $\widetilde{\Pi}^{21}_{xx}$ as
\begin{align}
    \widetilde{\Pi}^{21}_{xx}(\vec{q},i\Omega^{}_m)&=\int_{\vec{k},\omega^{}_n}\;\dfrac{2m\Omega^{}_m k_x^2}{(\omega_n^2+k^2+m^2)\qty[(\omega^{}_n+\Omega^{}_m)^2+(\vec{k}+\vec{q})^2+m^2]}
    \\&=\int_0^1d u\int_{\vec{k},\omega_n}\;\dfrac{2m\Omega^{}_m k_x^2}{[u(\omega^{}_n+\Omega^{}_m)^2+(1-u)\omega_n^2+k^2+u(1-u)q^2+m^2]^2}+\mathcal{O}(q^2),
\end{align}
where we have introduced Feynman parameters in the second line and shifted $\vec{k}\rightarrow \vec{k}-u\vec{q}$. Due to the Pauli matrix contractions, we have \begin{align}
    \widetilde{\Pi}^{31}_{y}(\vec{q},i\Omega^{}_m)&=-   \int_0^1d u\int_{\vec{k},\omega^{}_n}\;\dfrac{2\Omega^{}_m k_y^2}{[u(\omega^{}_n+\Omega^{}_m)^2+(1-u)\omega_n^2+k^2+u(1-u)q^2+m^2]^2}=-\dfrac{\widetilde{\Pi}^{21}_{xx}(\vec{q},i\Omega^{}_m)}{m}.
\end{align}
Continuing, we define $\Delta(u)=u(\omega_n+\Omega_m)^2+(1-u)\omega_n^2+u(1-u)q^2+m^2$, so that 
\begin{align}\label{eq:pitilde_21}
\tilde{\Pi}^{21}_{xx}(\vec{q},i\Omega^{}_m)&=2m\Omega^{}_m\int_0^1d u\int_{\omega^{}_n}\dfrac{d k^{}_xd k^{}_y}{4\pi^2}\;\dfrac{ k_x^2}{[k^2+\Delta]^2}.
\end{align}
Imposing a UV cutoff $\Lambda$ and taking the limit $T\rightarrow 0$, we obtain 
\begin{align}
\tilde{\Pi}^{21}_{xx}(\vec{q},i\Omega_m)&=\dfrac{m\Omega_m}{8\pi}\int_0^1d u\;\left(\Lambda-2\sqrt{u(1-u)q^2+m^2}\right)+\mathcal{O}(\Omega_m^2).
\end{align}
Now taking the $q\rightarrow 0$ limit, we get
\begin{align}
\tilde{\Pi}^{21}_{xx}(\vec{q},i\Omega^{}_m)=\dfrac{m\Omega^{}_m}{8\pi}\left(\Lambda-2|m|\right)
\end{align} 
so that, using Eq.~(\ref{eq:continuum_etaH_selfenergy}),
\begin{align} \Pi^{xy}(\vec{q},i\Omega^{}_m)&=\vec{q}^2(g^{}_1g^{}_2-g^{}_3g^{}_4)\dfrac{m\Omega^{}_m}{8\pi}\left(\Lambda-2|m|\right)+\vec{q}^2g^{}_0(g^{}_2-g^{}_3)(m+M)\dfrac{\Omega^{}_m}{8\pi}\left(\Lambda-2|m|\right)\\\implies\eta^H&=\dfrac{1}{4\pi L^2}\left[(g^{}_1g^{}_2-g^{}_3g^{}_4)m+g^{}_0(g^{}_2-g^{}_3)(m+M)\right](\Lambda-2|m|).
\label{eq:appendixeta_cont_zeroT}\end{align}
Now, to obtain the finite-temperature result, we go back to Eq.~(\ref{eq:pitilde_21}) to calculate
\begin{align}
\tilde{\Pi}^{21}_{xx}(\vec{q},i\Omega^{}_m)&=2m\Omega^{}_m\int_0^1d u\int_{\omega^{}_n}\dfrac{d k^{}_xd k^{}_y}{4\pi^2}\;\dfrac{k_x^2}{[k^2+\Delta]^2}
\\&=2m\Omega^{}_m\int_0^1d u\int\dfrac{d k^{}_xd k^{}_y}{4\pi^2}\;\dfrac{k_x^2}{4\xi^{3}_k}\cdot \qty[1-2n^{}_F(\xi^{}_k)+2\xi^{}_k n_F'(\xi^{}_k)]+\mathcal{O}(q,\Omega_m^2),
\end{align}
where we have evaluated the Matsubara sum, which is of the same form as for the lattice calculation, and defined $\xi_k\equiv \sqrt{k^2+m^2}$. Proceeding with the integral over $u$, we arrive at 
\begin{align}
\tilde{\Pi}^{21}_{xx}(\vec{q},i\Omega^{}_m)&=m\Omega^{}_m\int_0^\Lambda\dfrac{d k \cdot 2\pi k}{16\pi^2}\;\dfrac{k^2}{\xi^{3}_k}\cdot \qty[1-2n^{}_F(\xi^{}_k)+2\xi_kn_F'(\xi^{}_k)]\\
&=\dfrac{m\Omega^{}_m}{4\pi}\left(-|m|-2T\log\left(1+e^{-|m|/T}\right)\right)+\dfrac{m\Omega^{}_m}{8\pi}D^{}_\Lambda(m,T),
\end{align}
where we have defined the function 
\begin{align}
    D^{}_\Lambda(m,T)\equiv\frac{\Lambda ^2 \left(\frac{2}{e^{\frac{\sqrt{\Lambda ^2+m^2}}{T}}+1}-3\right)-2 m^2}{\sqrt{\Lambda ^2+m^2}}+4 T \log
   \left(e^{\frac{\sqrt{\Lambda ^2+m^2}}{T}}+1\right).
\end{align}
This brings us to
\begin{align} \Pi^{xy}(\vec{q},i\Omega^{}_m)&=\vec{q}^2\left[(g^{}_1g^{}_2-g^{}_3g^{}_4)m+g^{}_0(g^{}_2-g^{}_3)(m+M)\right]\dfrac{\Omega^{}_m}{8\pi}\cdot\left( D^{}_\Lambda(m,T)-2|m|-4T\log\left(1+e^{-|m|/T}\right)\right)\\
\implies\eta^H&=\dfrac{1}{4\pi L^2}\left[(g^{}_1g^{}_2-g^{}_3g^{}_4)m+g^{}_0(g^{}_2-g^{}_3)(m+M)\right]\cdot\left( D^{}_\Lambda(m,T)-2|m|-4T\log\left(1+e^{-|m|/T}\right)\right).
\label{eq:etah_continuum_appendix}
\end{align}
In the limit $m,T\ll\Lambda$, we have $D_\Lambda(m,T)\rightarrow \Lambda+4T\log\left(1+e^{-\Lambda/T}\right)$, and we can write the expression for $\eta^H$ in the continuum, at finite temperature, as
\begin{align} \eta^H=\dfrac{1}{4\pi L^2}\left[(g^{}_1g^{}_2-g^{}_3g^{}_4)m+g^{}_0(g^{}_2-g^{}_3)(m+M)\right]\cdot\left(\Lambda+4T\log\left(1+e^{-\Lambda/T}\right)-2|m|-4T\log\left(1+e^{-|m|/T}\right)\right),
\end{align}
\end{widetext}
the zero-temperature limit of which is in agreement with Eq.~\eqref{eq:appendixeta_cont_zeroT}. The finite-temperature continuum result allows us to rewrite the cutoff independent part of $\eta^H$ as
\begin{align}\eta^H&\sim |m|+2T\log\left({1+e^{-|m|/T}}\right)\nonumber
\\&= |m|+2T\log(2e^{-|m|/2T}\cosh(|m|/2T))\nonumber
\\   &= 2T\log(2\cosh(|m|/2T)),
\end{align} 
which leads to \equref{eq:etah_finiteT}. We observe that $\eta^H$ is analytic at all $T>0$ as $\cosh$ is an analytic and even function.

From Eq.~\eqref{eq:etah_continuum_appendix} above, we see that $\eta^H$ scales with the effective couplings $g_1g_2-g_3g_4$ and $g_0(g_2-g_3)$. This can be understood in the representation-theoretic framework. Writing out the $g_i$s in terms of irreducible representations as defined at the end of Sec.~\ref{sec:spinon_phonon_coupling_cont}, we find that both combinations
\begin{subequations}
\begin{alignat}{2}
    g^{}_1g^{}_2-g^{}_3g^{}_4&\propto g^{}_{A_1}g^{}_{A_2}+g^{}_{B_1}g^{}_{B_2},\\
     g^{}_0(g^{}_2-g^{}_3)&\propto g^{}_{A_1}g^{}_{A_2},
\end{alignat} 
\end{subequations}
transform under the $A_1\otimes A_2=B_1\otimes B_2=A_2$ representation of $C_{4v}$. Now, we expect $\eta^H$ to transform trivially under all symmetries (under $A_1$) as the phonon effective action must be invariant under all symmetries. This still holds true because $\eta^H$ can only exist in the presence of broken reflection symmetry, in which case, the symmetry of the phonon action is reduced $C_{4v}\rightarrow C_4$, and the $A_2$ of $C_{4v}$ descends to the trivial $A_1$ of $C_4$. As a result, $\eta^H$ has only one independent component with $C_4$ symmetry. More precisely, as the four-indexed Hall viscosity tensor is antisymmetric upon exchanging pairs of indices (phonon modes) while it is symmetric for exchange within each pair, we know that it has to transform under the antisymmetric $A_1$ tensor representation, which we denote $A_1^{a}$. Since the phonon field transforms under the vector representation $E_1$, this means that the independent component(s) of $\eta^H$ correspond to the component(s) of $A_1^a$ within $\bigwedge^2\mathrm{Sym}^2(E_1)$ (with $\mathrm{Sym}^2$ and $\bigwedge^2$ denoting the symmetrized and antisymmetrized tensor product, respectively). In our ansatz, we can illustrate this algebraically as
\begin{equation}
    {\textstyle \bigwedge }^2\mathrm{Sym}^2(E^{}_1)={\textstyle \bigwedge}^2(A^{}_1\oplus B^{}_1\oplus B^{}_2)=A_2^a\oplus B_1^a\oplus B_2^a,
\end{equation} 
in $C_{4v}$, which descends to $A_1^a\oplus 2B_1^a$ in $C_4$, so the Hall viscosity has one component. This procedure can also be carried out for other lattices. For example, as was shown for phonons with $C_{6v}$ symmetry \cite{yeperkins2020} on the honeycomb lattice, we have
\begin{equation}
    {\textstyle \bigwedge }^2\mathrm{Sym}^2(E^{}_1)={\textstyle \bigwedge}^2(A^{}_1\oplus E^{}_2)=A_2^a\oplus E_2^a,
\end{equation}
which descends to $A_1^a\oplus E_2^a$ in $C_6$, giving one independent component of $\eta^H$. As $A_2^a$ originated from $E_2\otimes E_2$, we know that $\eta^H$ must scale as $g_{E_2}^2$.

\bibliography{Refs.bib}

\begin{thebibliography}{92}
\bibitem [{dis()}]{dissipation_note}%
  \BibitemOpen
  \href@noop {} {}\bibinfo {note} {The rate of the work done by the system in
  terms of the stress tensor $T_{ij}$ is $\dot{W} = \int \dot{u}_i T_{ij}
  d{S}_j = \int \partial_j \dot{u}_i T_{ij} d{V}$, where the second equality
  follows from the divergence theorem. Since $T_{ij}$ is symmetric with respect
  to $(ij)$, it follows that $\partial_j \dot{u}_i T_{ij} = \eta^H_{ijkl}
  \dot{\epsilon}_{ij} \dot{\epsilon}_{kl}=0$, using the antisymmetric property
  $\eta^H_{ijkl} = -\eta^H_{klij}$.}\BibitemShut {Stop}%
\end{thebibliography}
\end{document}